# Glassy Dynamics of $LiCl.6H_2O$ Solution in Nanoporous Media

Armin Mozhdehei[1,*]

Mohammad Nadim Kamar[1]

Ronan Lefort[1]

Alain Moreac[1]

Arnaud Le Pottier[1]

Agathe Belime[2]

Denis Morineau[1]

1. Institute of Physics of Rennes, CNRS-University of Rennes, UMR 6251, F-35042 Rennes, France.
2. Institut Laue-Langevin, F-38042 Grenoble, France

* armin.mozhdehei@univ-rennes.fr

## Abstract

Understanding how nanoconfinement alters the dynamics of glass-forming aqueous electrolytes is essential for clarifying the interplay among ionic hydration, hydrogen-bond structure, and interfacial effects. Here, $LiCl.6H_2O$ was investigated in the bulk and under confinement in SBA-15 mesoporous silica with an average pore diameter of 8 nm. Differential scanning calorimetry, Raman spectroscopy, quasielastic neutron scattering, $^1$H spin–lattice relaxation, and pulsed-field-gradient NMR were combined to probe thermal behavior, hydrogen-bond structure, local mobility, and translational transport over complementary time and length scales. The calorimetric results show that $LiCl.6H_2O$ remains glass-forming under confinement, while its thermal signature of the glass transition becomes slightly broader and shifted upward relative to the bulk. Raman spectra in the O–H stretching region indicate that the concentrated LiCl solution possesses a weakened and less tetrahedrally connected hydrogen-bond network compared with bulk water. On the sub-nanosecond timescale, elastic fixed-window analysis reveals reduced mean-squared displacements under confinement, demonstrating suppressed motional amplitudes inside the pores. Inelastic fixed-window neutron scattering scans analyzed within a jump-diffusion framework yield lower effective translational diffusion coefficients and longer residence times for the confined liquid, indicating that confinement mainly hinders translational escape from transient local environments. $^{1}$H relaxometry further shows that confinement broadens the distribution of local proton fluctuation times, while PFG-NMR confirms that the measured long-range water mobility in bulk LiCl.6$H_2O$ solution is reduced relative to bulk water. While the present data do not resolve distinct interfacial and pore-centered populations in confined LiCl.6$H_2O$, its dynamics are markedly altered across timescales, from the glassy to the liquid state, resulting in slower, spatially constrained, and more heterogeneous motions.

## 1. Introduction

Water in the deeply supercooled regime exhibits unusual structural and dynamical behavior, but direct experimental access is severely hindered by rapid crystallization.[1-3] For this reason, highly concentrated aqueous lithium chloride solutions have long been used as model systems, because they strongly suppress ice formation and remain experimentally accessible over broad temperature ranges extending into the supercooled and glassy states.[4-7] Within this family, LiCl.6$H_2O$ is especially relevant because it is a robust glass former with a glass transition reported near 135–145 K and a well-established evolution from liquid to vitreous states.[1, 4, 8-11] Its selection is further supported by the LiCl–$H_2O$ phase diagram, which shows a V-shaped melting behavior, a change in the stable crystalline phase from ice Ih to salt hydrates with increasing salt concentration, and a pronounced no-crystallization region at intermediate-to-high LiCl contents.[5, 12-15] In the commonly used hydrate notation LiCl.$RH_2O$, the region of strongest glass-forming ability lies roughly around $4 < R < 6$, where crystallization is strongly suppressed. Accordingly, the choice of LiCl.6$H_2O$ in the present work was deliberate, because $R = 6$ lies inside the glass-forming region and is therefore particularly suitable for studying glassy and supercooled dynamics in both bulk and confined states.[4, 5]

However, concentrated LiCl solutions are not simply proxies for supercooled water, because the ions modify the liquid structure at the microscopic level.[1, 16] With increasing LiCl concentration, the extended hydrogen-bond network is progressively weakened, while the local arrangement of molecules becomes increasingly governed by ionic hydration.[1, 17] At the same time, LiCl.6$H_2O$ retains important water-like structural motifs: diffraction studies have shown that its glass and supercooled-liquid states remain relatively similar, whereas a much more pronounced reorganization occurs on heating into the normal liquid near 207 K.[18, 19] In particular, the O–O correlation near 4.4 Å, usually associated with tetrahedral short-range order, is absent in the high-temperature liquid but reappears in the glassy and supercooled states, showing that cooling restores a more correlated low-temperature structure despite the very high salt concentration.[18] This structural evolution is accompanied by equally rich dynamical behavior.

Quasielastic neutron scattering studies have shown that LiCl.6$D_2O$ exhibits stretched-exponential structural relaxation and a change from Arrhenius to super-Arrhenius temperature dependence on cooling, highlighting the complex dynamical behavior of this glass-forming solution.[1, 4, 20] In parallel, low-frequency Raman and inelastic neutron scattering studies showed that anharmonic contributions to the vibrational dynamics are negligible in the vitreous state but increase markedly in the supercooled state, while the glass remains structurally compact.[4] In this way of reasoning, these results show that LiCl.6$H_2O$ displays dynamical behavior spanning a broad range of processes, from localized vibrational motions to slower collective relaxation.

NMR studies provide an important complementary perspective because they probe both local molecular motion and long-range translational diffusion.[16, 21] In glass-forming LiCl water mixtures, NMR measurements have shown that water and lithium dynamics remain strongly coupled in the weakly supercooled regime, whereas different observables begin to decouple upon deeper cooling.[21, 22] Although the present measurements concern LiCl.6$H_2O$, earlier NMR work on LiCl.7$H_2O$ remains directly relevant because LiCl–$H_2O$ solutions in the range $3 < R < 7$ belong to the same glass-forming concentration regime, where the deeply supercooled dynamics depend only weakly on nominal composition.[21] At eutectic composition ($R = 7$), proton NMR line-shape analysis further suggested the coexistence of distinct local water environments and a crossover near 225 K associated with increasing tetrahedral character at low temperature.[2] Together, these studies establish that concentrated LiCl solutions already display complex multiscale dynamics in the bulk state, providing an essential reference frame for assessing the additional effects of nanoconfinement.[2, 21]

Once transferred into a mesoporous host, the liquid becomes subject to additional constraints arising from pore geometry, surface chemistry, and interfacial interactions.[5, 23] Mesoporous silica is particularly relevant in this context because its pore walls are terminated by silanol groups, which can influence the confined electrolyte through hydration-mediated ion–surface interactions in addition to purely geometric restriction.[16, 24, 25] Previous neutron and NMR studies on LiCl aqueous solutions confined in silica matrices, including MCM-41 and SBA-15, showed that confinement significantly modifies both local and diffusive dynamics, with the magnitude of the effect depending on temperature, pore size, and the nature of the confined electrolyte.[16, 23, 26] These observations make mesoporous silica an appropriate model system for examining how nanoconfinement modifies the dynamical behavior of a glass-forming concentrated electrolyte.

Earlier calorimetric studies on LiCl aqueous solutions confined in mesoporous silica likewise showed that nanoconfinement modifies the thermal behavior of these systems while preserving their glass-forming character.[5, 27] More generally, phase transitions of aqueous salt solutions in nanopores are known to be sensitive to pore size, interfacial effects, and confinement-induced changes in the liquid environment.[27, 28]

These observations motivate deeper exploration, particularly into how water motions are influenced across scales ranging from pore dimension down to molecular size. Neutron scattering is particularly valuable because it probes molecular motion on the nanometer and nanosecond scales that are directly relevant to local mobility in confined liquids.[1, 9] In particular, high-resolution backscattering experiments provide access to the temperature dependence of the elastic and inelastic response within well-defined instrumental time windows.[29] This makes fixed-window analysis especially suitable for comparing the dynamical evolution of bulk and confined LiCl.6$H_2O$ and for identifying the temperature range over which confinement modifies the onset of molecular mobility.

In this study, we examine the dynamics of LiCl.6$H_2O$ in the bulk and under confinement in SBA-15 mesoporous silica with an average pore diameter of 8 nm by combining differential scanning calorimetry, $^1$H NMR relaxometry, pulsed-field-gradient NMR, Raman spectroscopy, and elastic and inelastic fixed-window neutron scattering. This multi-technique approach is designed to connect thermal behavior, hydrogen-bond structure, local correlation times, long-range self-diffusion, and short-time neutron-scale mobility within a unified picture of a glass-forming concentrated electrolyte under nanoconfinement. More specifically, we aim to determine how confinement in SBA-15 modifies the thermal response, motional amplitudes, translational mobility, and glassy dynamics of LiCl.6$H_2O$ relative to the bulk.

# 2. Materials and methods

## 2.1. Samples

Lithium chloride (LiCl; ≥99.9% purity) was purchased from Sigma-Aldrich (Merck) and used as received. Ultrapure deionized water (18.2 MΩ·cm at 21.8 °C) was produced on an ELGA PURELAB Classic DI system. SBA-15 mesoporous silica was synthesized by a surfactant-templated route analogous to prior works,[30-34] using the nonionic triblock copolymer Pluronic P123, $(EO)_{20}(PO)_{70}(EO)_{20}$, as a structure-directing agent, yielding a hexagonally ordered array of aligned cylindrical mesopores. Pore characteristics were determined from $N_2$ adsorption isotherms. Analysis by the KJS method gave an average mesopore diameter of approximately $D_p \approx 8$ nm, whereas HK analysis revealed an additional microporous contribution centered at about 1 nm. The total pore volume was in the range of $\rho \approx 0.81 – 0.94$ cm$^3$/g.

Lithium chloride hexahydrate, LiCl.6$H_2O$, was deliberately selected because it lies in the glass-forming region of the LiCl–$H_2O$ phase diagram, where crystallization is strongly suppressed, and the system can be studied in the supercooled and glassy states.[5] Sample preparation was similar for all the confined samples. SBA-15 mesoporous silica was first calcined and then dried under primary vacuum at 120 °C for 18 h to remove residual adsorbates. The dried matrix was subsequently filled by liquid imbibition using LiCl.6$H_2O$, while a bulk sample of identical composition was prepared as the reference. The amount of liquid introduced was adjusted to about 0.9 cm$^3$/g, corresponding to the 100% accessible pore volume of SBA-15 estimated from water adsorption, so as to achieve complete pore filling while avoiding excess liquid outside the pores. After loading, the samples were allowed to equilibrate for 24 hours by placing them in a closed desiccator at a relative humidity fixed by a LiCl.6$H_2O$ solution to ensure homogeneous distribution of the liquid within the porous network.

## 2.2. Differential Scanning Calorimetry (DSC)

Differential scanning calorimetry (DSC) was used to determine the glass-transition temperature ($T_g$) of the investigated samples. Measurements were carried out on a TA Instruments Q20 calorimeter equipped with liquid-nitrogen cooling, using the same procedure as in our earlier studies.[33, 35, 36] The calorimeter was calibrated in both temperature and heat flow with indium.

Samples were sealed in hermetic TZero© aluminum pans, with a matched empty pan serving as the reference, and all experiments were performed under a dry helium atmosphere. In each DSC cycle, the sample was maintained at 293 K for 1 min, cooled to 125 K at 10 $K.min^{-1}$, held isothermally for 1 min, and then reheated to 293 K at 10 $K.min^{-1}$. Heat-flow thermograms were processed with TA Instruments Universal Analysis software. Transition onset temperatures were determined by the tangent-intersection method, and peak temperatures were assigned to the maxima in the heat-flow signal.

## 2.3. Quasi-elastic neutron scattering (QENS) measurements

Neutron scattering measurements were carried out at the Institut Laue-Langevin (ILL, Grenoble, France) on the high-resolution backscattering spectrometer IN13 with the wide *Q*-range (up to 5 $Å^{-1}$). The incident wavelength was set to 2.23 Å, leading to an energy resolution of 8 μeV (FWHM). The sample transmission was approximately 92%. The standard data reduction was performed using standard ILL programs (LAMP (Large Array Manipulation Program)), which correct for incident flux, cell scattering and self-shielding using an angle-dependent slab correction. To compensate for detector efficiency, the data were normalized to either a vanadium or a spectrum taken at 20 K.

The experiment was performed in the fixed-window mode in order to monitor the temperature evolution of the dynamics of bulk $LiCl.6H_2O$ and $LiCl.6H_2O$ confined in SBA-15 over the temperature range from 300 to 20 K. In neutron fixed-window experiments, the measured incoherent intensity $I(Q, \omega, T)$ is recorded as a function of momentum transfer $Q$, energy transfer $\hbar\omega$, and temperature $T$. The accessible dynamical information is determined by the instrumental energy resolution and by the selected energy offset. As shown in Fig. 1, in the present work, two complementary fixed-window acquisition modes were used on IN13: an elastic fixed-window scan (EFWS) measured at zero energy transfer, $\Delta E = 0$ μeV, and an inelastic fixed-window scan (IFWS) measured at a finite energy offset of $\Delta E_{offset} = 26$ μeV.

The EFWS records the intensity within the instrumental elastic window around $\omega = 0$. Accordingly, the elastic intensity increases when the molecular motions (e.g., translation, rotation) slow down upon cooling, when their corresponding quasielastic broadening becomes smaller than the instrumental resolution $\delta E_{res} = 8$ μeV. This corresponds to dynamics with typical relaxation times longer than ($\tau_{res} = \hbar/\ \delta E_{res}$) about 0.08 ns. The elastic intensity also varies on cooling with the increase of the Debye-Waller factor associated with vibrational motions that prevail in the glassy state. The EFWS therefore provides a direct way to monitor the gradual dynamical slowdown of the system and to determine the temperature dependence of the mean-squared displacement (MSD). The IFWS, by contrast, records the intensity at a finite energy transfer and is hence sensitive to relaxational processes whose characteristic time scale becomes comparable to the time scale associated with the chosen offset energy. As a result, the IFWS is particularly useful for identifying the temperature region where a given dynamical process crosses the instrumental window. In the present study, the IFWS data were used to extract the characteristic quasielastic linewidth $\Gamma_T$ and, from its $Q$-dependence, effective translational diffusion parameters.

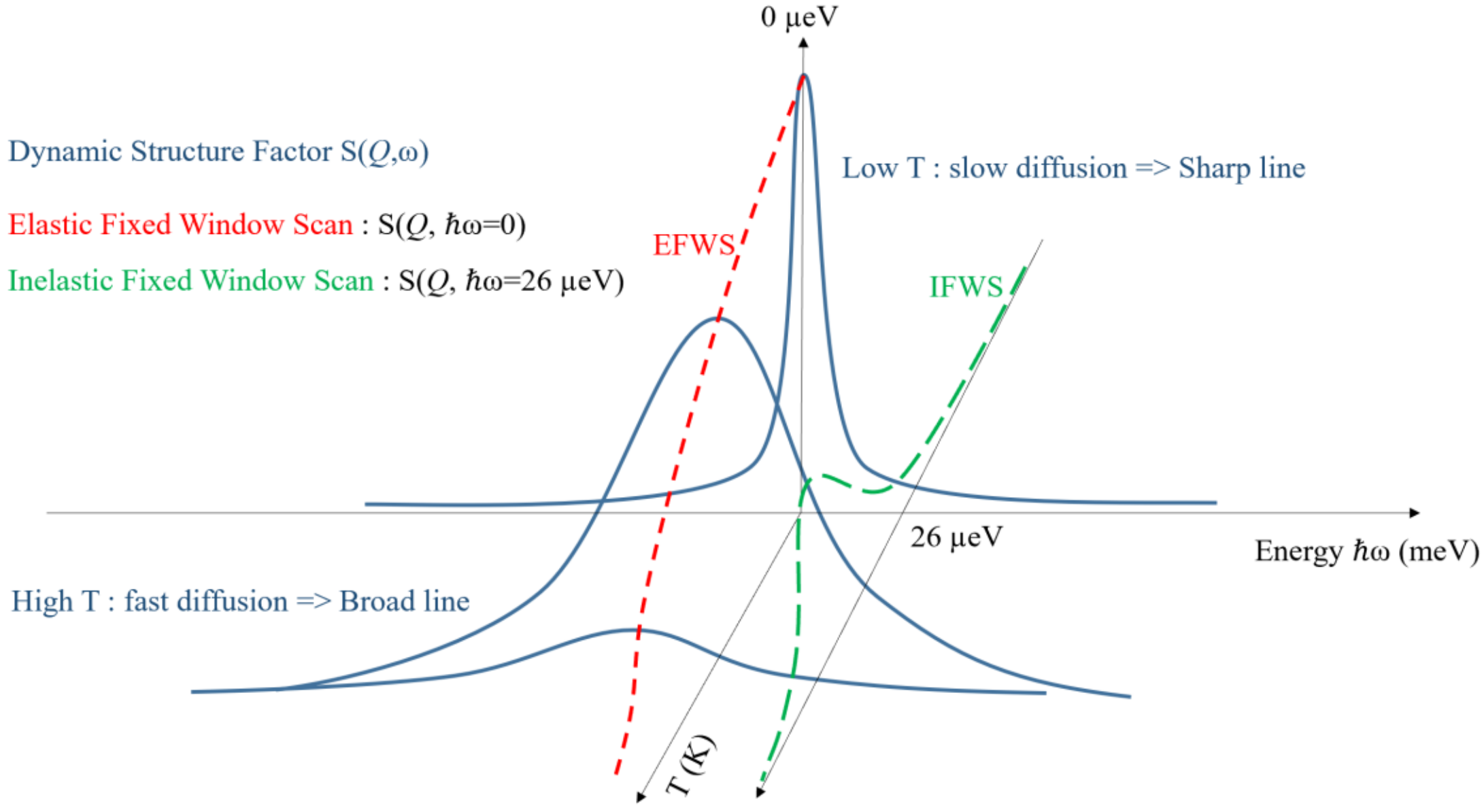

*Fig. 1. Schematic view of the fixed window scan (FWS) method on the IN13 instrument at ILL.*

Data were collected for the bulk LiCl.6$H_2O$ sample, the LiCl.6$H_2O$-filled SBA-15 sample, and the dry SBA-15 matrix, contained in hermetically closed rectangular aluminum cells. By measuring the dry SBA-15 matrix, we identified and subtracted the scattering contribution of the silica host from the filled one, thereby isolating the signal of the confined aqueous phase. Prior to analysis, all neutron data were corrected for background and empty-cell scattering. Due to the large incoherent cross section of hydrogen, the resulting intensity refers to the self-part of the dynamic structure factor of water molecules in the aqueous solutions.

### 2.4. NMR measurements: PFG-NMR diffusion and relaxometry

All NMR measurements were carried out on a Bruker DMX 300 spectrometer fitted with a diffusion probe capable of delivering magnetic-field gradient pulses up to 30 $T.m^{-1}$ along the vertical $z$ direction. The experiments were performed over the temperature range 200–350 K, with temperature regulated by a Bruker BVT 3000 variable-temperature controller. Samples were contained in 5 mm glass NMR tubes filled to an effective height of about 20 mm and were held at each set temperature for 20 min prior to acquisition to ensure thermal equilibration, similar to that used previously.[37] The recorded spectra were subsequently processed and analyzed using Bruker TopSpin 4.4.1.

Translational self-diffusion coefficients were determined using a pulsed-field-gradient stimulated-echo (PGSTE) sequence. The PGSTE sequence was preferred over the conventional pulsed-field-gradient spin-echo (PGSE) sequence because it provides improved signal-to-noise performance in systems where transverse relaxation ($T_2$) is relatively fast. The diffusion-induced attenuation of the NMR signal, $E$, was analyzed using the standard Stejskal-Tanner formalism, in which the measured echo attenuation is related to the self-diffusion coefficient $D$, the gyromagnetic ratio $\gamma_H$, the gradient pulse strength $g$, the gradient duration $\delta$, and the observation time $\Delta$ (i.e., $E = exp\left[-D\gamma_H^2 g^2 \delta^2 (\Delta - \frac{\delta}{3})\right]$). In the present measurements, the observation time was fixed at 20 ms, while the gradient amplitude was varied in 16 increments.

For the bulk sample, the attenuation of the NMR signal was well described by a single-exponential decay, yielding a long-range self-diffusion coefficient on the micrometer scale. For the confined sample, however, the measured attenuation should be interpreted more cautiously. On the PFG-NMR length scale, both the encoding length $q^{-1}$ and the characteristic displacement during the diffusion delay are much larger than the SBA-15 pore diameter. Under these conditions, the attenuation can be influenced by restricted motion in the pore geometry, powder averaging over pore orientations, and exchange between confinement environments. Therefore, the confined PFG-NMR values discussed in this work are treated as effective long-range transport parameters rather than as unique unrestricted intrapore self-diffusion coefficients. This distinction is important when comparing with neutron fixed-window measurements, since PFG-NMR probes translational motion over much longer time and length scales than the IN13 neutron backscattering spectrometer.

It should be noted that the approximate $q$-range covered for the confined system extends from about 0.35 to 8 $\mu m^{-1}$, whereas for the bulk it spans only about 0.02 to 0.42 $\mu m^{-1}$. This difference is already important at a qualitative level, because it shows that the confined and bulk NMR measurements do not probe the same spatial window. The corresponding encoding lengths $q^{-1}$ are about 0.13 to 2.9 μm for the confined sample, which are orders of magnitude larger than the 8 nm SBA-15 pore diameter.

In addition to diffusion measurements, proton spin-lattice ($T_1$) was measured in order to characterize the local dynamical behavior of $LiCl.6H_2O$ in the bulk and confined states. The spin-lattice relaxation time $T_1$ probes the return of the nuclear magnetization to thermal equilibrium along the external magnetic field and is sensitive to molecular motions with correlation times close to the inverse Larmor frequency. $T_1$ was used here as a probe of local mobility and spectral-density changes, whereas PFG-NMR was used to access long-range translational diffusion. The relaxation data were analyzed as a function of temperature in order to compare the dynamical evolution of water molecules in the bulk $LiCl.6H_2O$ and $LiCl.6H_2O$ confined in SBA-15. In this way, the NMR measurements provide complementary access to both the microscopic correlation times and the macroscopic transport properties of the system.

## 2.5. Raman spectroscopy

Raman spectra were recorded in the O–H stretching region in order to probe the hydrogen-bond network of bulk and confined water and $LiCl.6H_2O$. Measurements were performed in the 2600–3900 $cm^{-1}$ spectral range at $T = 293$ K using a LabRAM HR800 Raman spectrometer (Horiba), following an approach similar to that used previously for confined water in mesoporous silicas.[38] A 532 nm laser line was used as the excitation source, and the incident beam was focused onto the sample surface through a long-working-distance Olympus 50× ULWD objective. The scattered light was collected in backscattering geometry, and the Rayleigh contribution was removed by dielectric edge filters. To facilitate direct comparison between spectra, all Raman spectra were normalized to the maximum intensity of the main O–H stretching band.

# 3. Results and Discussion

## 3.1. Differential Scanning Calorimetry

We begin by examining the DSC results, which provide the thermal framework for identifying how confinement in SBA-15 modifies the $T_g$ of LiCl.6$H_2O$ relative to the bulk. Fig. 2 compares the DSC thermograms of bulk LiCl.6$H_2O$ with that of LiCl.6$H_2O$ confined in SBA-15. For both systems, the absence of exo/endothermic peaks confirms that no crystallization or melting events occur over the scanned $T$ range. This is consistent with the deliberate choice of the $R$ = 6 composition inside the glass-forming region of the LiCl–$H_2O$ phase diagram.[1, 4, 8-11] For the bulk sample, the glass transition is identified from the onset of the heat-capacity step, which occurs near $T_{g,onset}$ = 141 K. This assignment is consistent with earlier work that placed the glass transition of LiCl.6$H_2O$ in the approximate range 130–145 K.[4] The confined SBA-15 sample exhibits a similar heat capacity jump, but the signal is broader and the onset is shifted to slightly higher temperature than in the bulk ($T_{g,onset}$ = 144 K). This indicates that confinement does not suppress the glass transition, but modifies its thermal signature. Such behavior is in good agreement with calorimetric studies of confined LiCl aqueous solutions, where vitrification persists under silica confinement and the onset of the glass-transition interval can shift modestly upward for stronger confinement.[5, 27] This upward shift and broadening of the glass transition are commonly attributed to the dynamical slowdown and increased dynamic heterogeneity induced by confinement.[30] It is worth mentioning that in the present work these DSC data mainly provide the thermal reference frame for the neutron and NMR results rather than constituting the primary source of novelty.

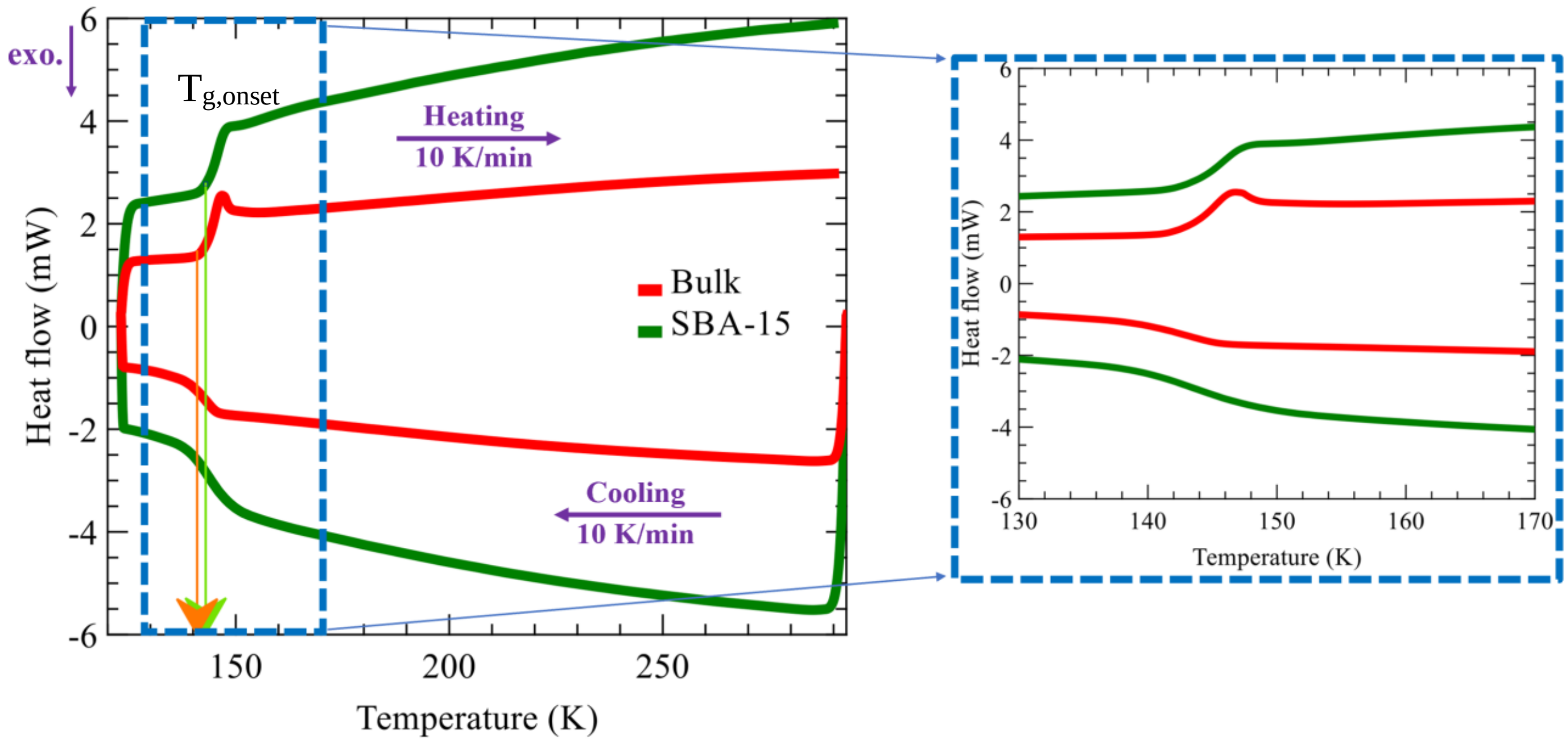

*Fig. 2. DSC thermograms of bulk LiCl.6$H_2O$ and confined in SBA-15 (8 nm pores). Cooling: 10 K.min$^{-1}$ to 125 K; heating: 10 K.min$^{-1}$ (exothermic down). Heating and cooling traces are shown for the bulk sample (red) and the confined sample (green). The low-temperature anomaly in the vicinity of 140–145 K differs in position and shape for the two systems, indicating that nanoconfinement modifies the thermal evolution of LiCl.6$H_2O$. The arrows mark the onset temperatures of the glass-transition interval $T_{g,onset}$ discussed in the text.*

## 3.2. Elastic Fixed Window Scans: mean squared displacement

We next turn to the IN13 fixed window scan neutron-scattering results, which probe the temperature dependence of local mobility in bulk and confined LiCl.6$H_2O$ on the nanosecond timescale. Fig. 3 presents the elastic fixed window scan intensity of the bulk sample, the dry silica matrix, and the confined LiCl.6$H_2O$ system. It provides the first direct neutron-scattering evidence that the molecular dynamics of the solution differ between bulk and confined environment on the sub-nanosecond time scale. The EFWS intensity summed over the measured $Q$ range of bulk LiCl.6$H_2O$ decreases strongly on heating from low temperature (20 K) to room temperature (300 K), which is the expected behavior when progressively more molecular motions become fast enough to transfer spectral weight out of the elastic line. At temperatures lower than 150 K, i.e., in the glassy state, most relaxational modes are too slow to be resolved from the elastic contribution. Consequently, the EFWS is dominated by the Debye–Waller-like term (i.e., vibrational motion) and remains large. As the temperature rises above the glass transition, the onset of anharmonic and diffusive motions, marked by quasielastic broadening, results in a more pronounced reduction of the elastic intensity.

The dry SBA-15 matrix exhibits only a weak and nearly featureless temperature dependence over the whole scanned range, showing that the strong temperature-dependent signal observed for the filled sample arises mainly from the confined $LiCl.6H_2O$ phase rather than from the silica host. For direct comparison between bulk and confined $LiCl.6H_2O$, the intensity of the EFWS was rescaled by their minimum and maximum values (corresponding to 20 K and 300 K, respectively) and illustrated in the inset of Fig. 3. A main observation is that the intensity of the confined sample is shifted upward by about 10 K. This is also consistent with the observed increase in the glass transition calorimetric signature discussed in the previous part.

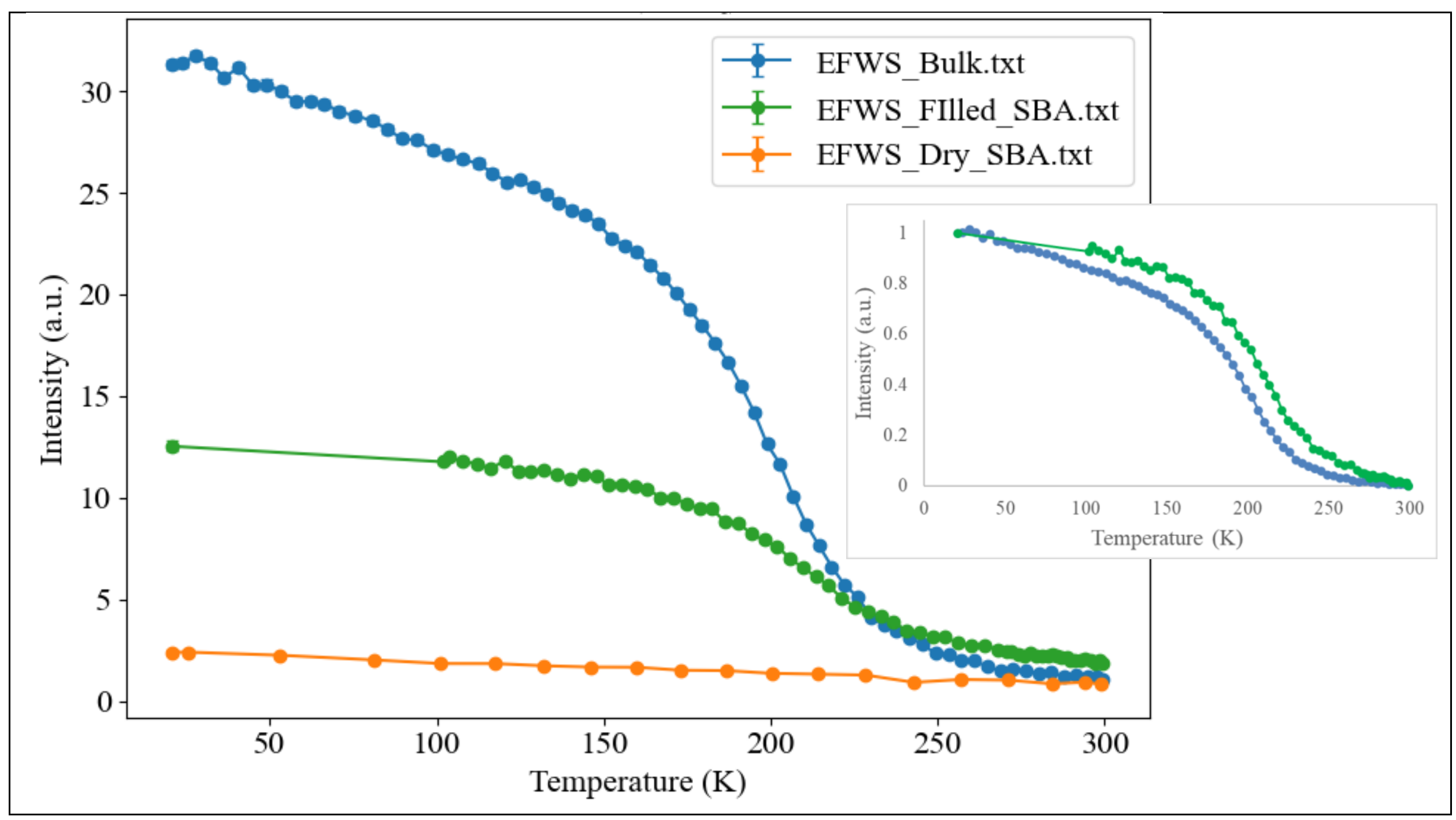

*Fig. 3. Temperature dependence of the elastic fixed-window scan (EFWS) intensity summed over the full measured Q range for bulk $LiCl.6H_2O$, dry SBA-15, and SBA-15 filled with $LiCl.6H_2O$. Inset: EFWS with intensity rescaled between zero and unity.*

To go beyond the EFWS intensities and quantify the amplitude of motion on the IN13 timescale, the *Q*-dependence of the EFWS signal was analyzed using the Gaussian approximation, as previously applied in related confined-liquid studies.[39] This Gaussian approximation is valid when the quasielastic broadening remains sufficiently narrow compared with the instrumental resolution and the dominant contribution to the elastic signal arises from localized displacements. In that limit, the EFWS intensity can be written as

$$I_{EFWS}(Q) \propto \exp\left(-\frac{\langle r^2\rangle Q^2}{6}\right) \quad (1)$$

where $\langle r^2\rangle$ is the mean-squared displacement (i.e., twice the value of $\langle u^2\rangle$, the mean squared offset of atomic positions from their equilibrium positions).

This expression is the fixed-window analogue of the Debye–Waller treatment and is particularly appropriate in the low-temperature regime, where the system is still dominated by glassy or weakly anharmonic motions. Under these conditions, a plot of $\ln I_{EFWS}(Q)\ vs.\ Q^2$ is expected to be close to linear, and the slope provides direct access to the MSD. At higher temperature, however, the EFWS no longer follows a simple exponential decay with $Q^2$. The signal becomes affected by the onset of quasielastic processes and by increasing anharmonicity, which leads to noticeable curvature in the $Q^2$-representation. In order to account for this behavior, fitting of a more general expression proposed by Zorn[40] is provided in the Supporting Information.

The MSD was extracted from the $Q$-dependence of the EFWS intensity using the Gaussian approximation; representative fits for bulk and confined $LiCl.6H_2O$ are provided in the middle panels of Fig. S1. The MSD values extracted from these fits are shown in Fig. 4. In both systems, the MSD increases gradually at low temperature and then rises more rapidly on heating through the glass-transition region. Such behavior reflects the crossover from predominantly vibrational or weakly anharmonic motion in the glassy state to the activation of larger-amplitude localized displacements, such as librational, cage-rattling, or short-range diffusive displacements, at higher temperature. For bulk $LiCl.6H_2O$, the MSD remains small and nearly linear up to about 145–150 K, i.e., close to $T_{g,onset}$, and then increases sharply. The confined sample follows the same qualitative trend, but the increase is shifted upward and systematically weaker, so that the absolute MSD remains lower over the entire temperature range. The smaller MSD of the confined sample below $T_{g,onset}$ indicates that the liquid explores a smaller local configurational space on the IN13 timescale. This interpretation is consistent with simulation studies of confined liquids, in which reduced short-time cage-rattling amplitudes near the interface were linked to a stiffer local environment and to a confinement-imposed static energy landscape.[41, 42] In this sense, the lower

MSD in SBA-15 is consistent with stronger localization of short-time motion in the pore environment.

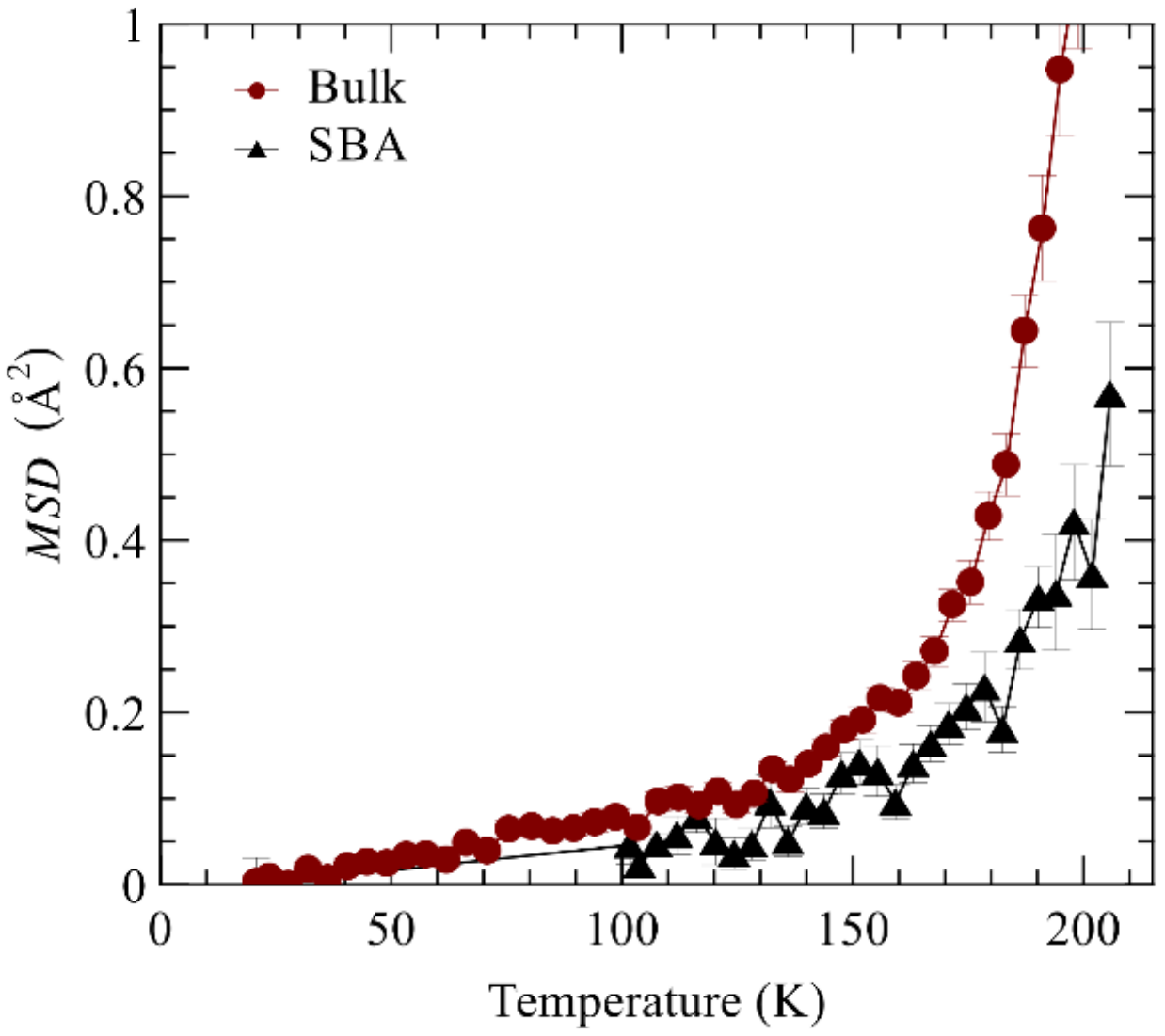

*Fig. 4. Temperature dependence of the mean-squared displacement (MSD) extracted from the Gaussian analysis for bulk LiCl.6$H_2O$ (circles) and LiCl.6$H_2O$ confined in SBA-15 (triangles).*

### 3.3. Inelastic Fixed Window Scans: translational motion

The IFWS provides complementary information to the EFWS because it follows the redistribution of spectral weight at a finite energy offset rather than at the elastic line.[43] In the present experiment, the IFWS was measured at a single offset energy of $E_{offset}$ = 26 μeV, so it is particularly sensitive to molecular motions whose characteristic timescale becomes comparable to the Fourier timescale associated with this energy transfer, namely $\tau = \hbar / E_{offset} \approx 25$ ps. While EFWS primarily provides insights into vibrational dynamics in the glassy state and highly viscous liquids near $T_{g,onset}$, IFWS is particularly well-suited for investigating translational motions at higher temperature in the liquid state.

Fig. 5 shows that the IFWS intensity summed over the full measured $Q$ range for both bulk and confined LiCl.6$H_2O$ increases on heating, reaches a broad maximum in the intermediate temperature range, and then decreases again at higher temperatures. This is the characteristic signature expected when a relaxational process progressively enters, then exits, the dynamic window selected by the fixed inelastic offset. As shown in the middle panel of Fig. 5, at low temperature, the translational quasielastic linewidth $\Gamma_T(T)$ is narrower than the selected offset, and only a limited fraction of the quasielastic intensity is transferred into the inelastic window. As temperature increases, the corresponding relaxation becomes faster and the linewidth approaches the offset energy, which leads to an increase in the IFWS intensity. Once the linewidth exceeds that energy, the inelastic signal at the fixed offset decreases again because the spectral weight spreads over a wider energy range. This explains why the IFWS exhibits a broad maximum peak rather than a monotonic evolution with temperature. The position of this maximum marks the temperature interval where the translational dynamics is best matched to the experimental timescale probed by the 26 μeV energy window.

The dry SBA-15 matrix exhibits an almost constant signal across the entire temperature range, confirming that the porous host behaves as a rigid solid material with no detectable diffusion process. Hence, the observed IFWS variation of the filled sample is dominated by the confined LiCl solution rather than by the silica host. A comparison between bulk and confined LiCl.6$H_2O$ shows that both systems display the same general IFWS behavior, but the detailed shape and intensity differ. The confined sample exhibits a slightly weaker modulation and is shifted to the high temperature side of the peak, indicating that the translational dynamics is reduced in confinement. This is fully consistent with the temperature evolution of the EFWS discussed earlier and supports reduced dynamics in SBA-15.

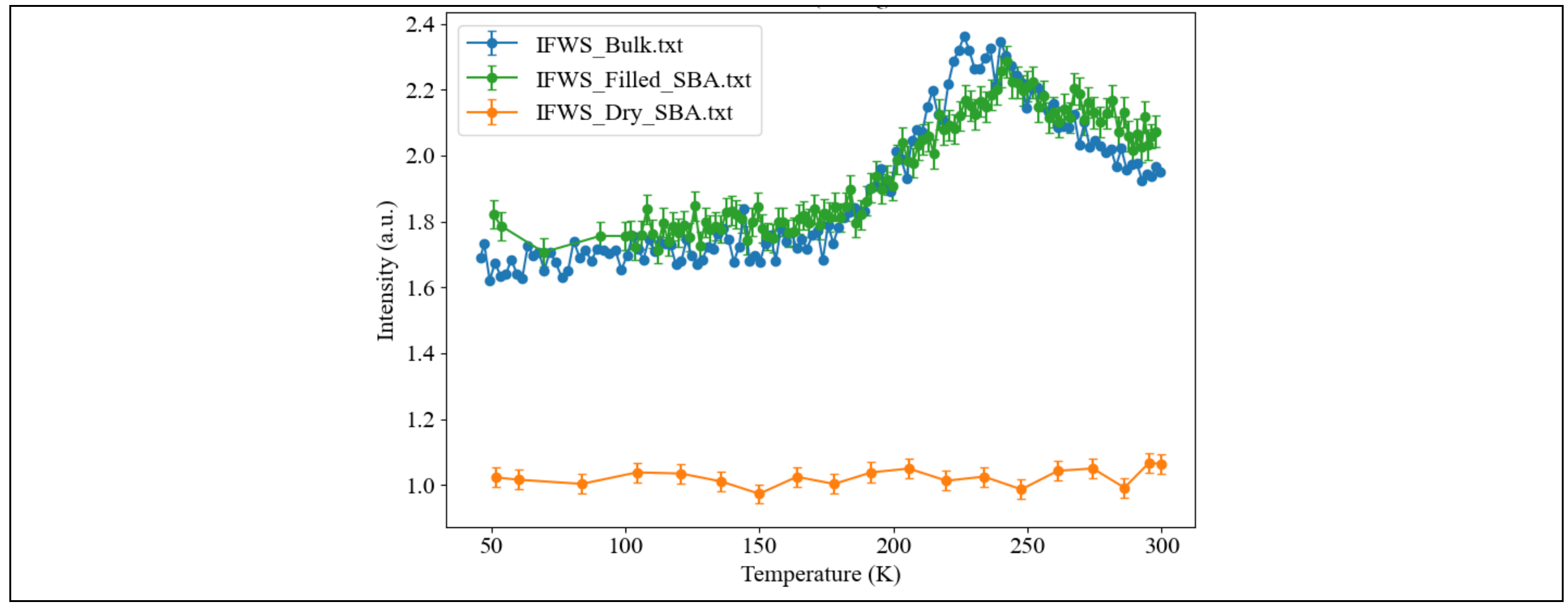

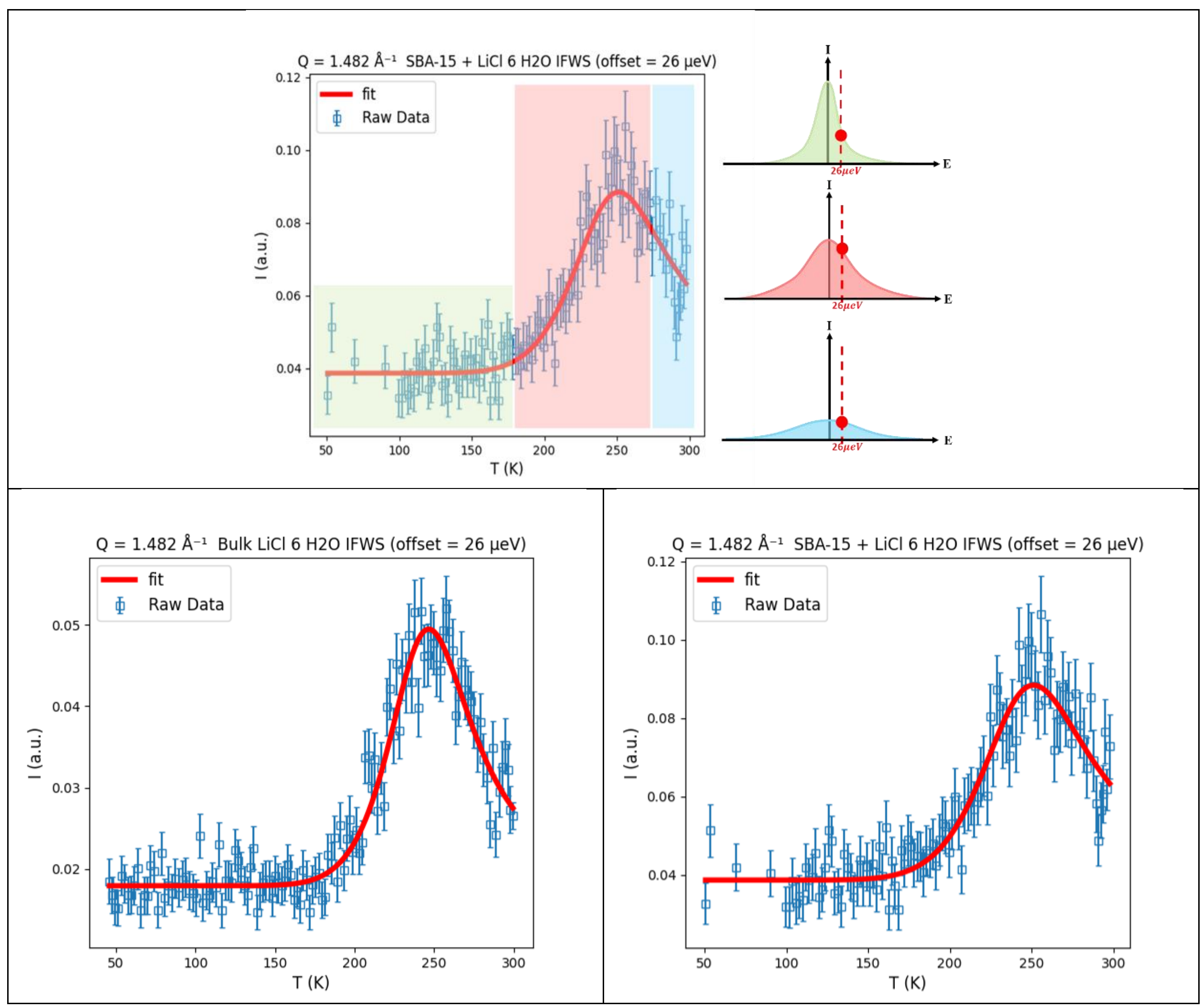

*Fig. 5. Top panel: temperature dependence of the IFWS intensity summed over the full measured Q range for bulk LiCl.6$H_2O$, dry SBA-15, and SBA-15 filled with LiCl.6$H_2O$, measured at an energy offset of 26 μeV. Middle panel: showing the variation of the half width at half maximum at the defined energy offset and its variation by temperature. Bottom panels: representative fits of the IFWS temperature dependence at selected Q values (as representation of all Qs) for bulk LiCl.6$H_2O$ (left) and LiCl.6$H_2O$ confined in SBA-15 (right). The red lines represent the fitted model used to extract the temperature-dependent translational linewidth $\Gamma_T(T)$.*

In order to interpret the dynamics probed by the IFWS, we refer to the classical description used for liquid water. There is broad consensus that water dynamics, as probed by neutron scattering, can be decomposed into three contributions: vibrations, local motions, and translational diffusion. This classical description is supported by observations that the dynamic structure factor exhibits two quasielastic components across multiple studies.[44-46] While the narrow component is widely accepted as diffusive, the nature of the broad component remains debated. Teixeira et al. initially attributed it to continuous rotational diffusion,[44] but later reinterpretations proposed it as picosecond structural fluctuations within dynamical basins[45] or non-diffusive reorientation involving large angular jumps.[47]

In contrast, the narrow component is generally interpreted as translational jump-diffusion, which is discontinuous at the molecular scale and progressively deviates from Fick's law at high $Q$. Furthermore, all four cited studies, and others, demonstrate that the quasielastic spectra align with these two components being Lorentzian (Debye-type). Alternative models, such as a single stretched exponential (inspired by the relaxing cage model and the heterogeneous dynamics of glass-forming liquids), have also been applied to water. However, it has been argued that the apparent fit of stretched exponential functions to QENS spectra stems from the inability to distinguish fast local dynamics from longer-ranged diffusion.[45]

In our case, the experimental data from high-resolution inelastic fixed-window scans do not cover the full spectrum over a broad energy range (typically ±1 meV), which is necessary to separate the two quasielastic components.[45, 46] Within the much narrower energy range of our finite energy offset ($\Delta E_{\mathrm{offset}}$ = 26 μeV), we infer that the QENS intensity is dominated by the narrow, slowest diffusive component.

To quantify the IFWS behavior, each temperature-dependent IFWS curve was analyzed at fixed $Q$ using the same approach as in the reference analysis.[37] The model assumes that the dynamic structure factor at each $Q$-value can be approximated by a sum of two contributions: - A Lorentzian, $L_T(Q,\omega)$ with amplitude $A(Q)$ and width $\Gamma_T(Q,T)$ associated with translational diffusion; - A flat background $BG(Q)$ representing broader quasielastic components arising from local fast motions that exceed the instrument's dynamical range. This model was evaluated at the specific value $\hbar\omega_{\mathrm{offset}}$ = 26 μeV, leading to function $I_{IFWS}^{Fit}$ (Eq. 2), which was fitted to the experimental IFWS to extract the three parameters $A(Q)$, $BG(Q)$ and $\Gamma_T(Q,T)$.

$$I_{IFWS}^{Fit} = A(Q)L_T\big(Q,\omega_{offset}\big) + BG(Q) \tag{2}$$

All three parameters of $I_{IFWS}^{Fit}$ were fitted independently for each $Q$-value, but the temperature dependence of the Lorentzian width $\Gamma_T(Q,T)$ (HWHM) was constrained to follow an Arrhenius behavior with a unique, $Q$-independent, activation energy ($E_a$), according to:

$$\Gamma_T(Q,T) = \Gamma_T^0(Q)\exp\left(-\frac{E_a}{k_BT}\right) \tag{3}$$

The representative fits shown in the bottom panel of Fig. 5 reproduce the data well, indicating that the observed inelastic response can indeed be interpreted in terms of a thermally activated translational process crossing the fixed energy window. In fact, the prefactor $\Gamma_T^0(Q)$ and activation energy $E_a$ obtained from fitting the temperature evolution of the intensities represent the Arrhenius dependence of the translational linewidth $\Gamma_T(Q,T)$ that best captures the full evolution of the IFWS signal over the entire measured temperature range. Although the fitted Arrhenius form describes the whole temperature dependence, the most reliable determination of $\Gamma_T$ is obtained near the peak region of the IFWS curve. From a direct observation of the $Q$-dependence of the peak position [Fig. S3] we deduce that this temperature range lies between 200 K and 250 K. Over this relatively narrow temperature range, the values of $\Gamma_T(Q,T)$ are not expected to vary by more than a decade. This accounts for the Arrhenius model's ability to accurately reproduce the experimental data, though deviations may emerge over a broader timescale. Conversely, efforts to incorporate non-Arrhenius temperature dependence into the fitted function yielded unreliable parameters.

The $Q$-dependence of $\Gamma_T(Q,T)$ is illustrated in Fig. 6 at three selected temperatures. At low $Q$, $\Gamma_T$ grows approximately linearly with $Q^2$ for both bulk and confined samples. This low-$Q$ regime corresponds to the continuous Fickian translational diffusion. At larger $Q$, however, the linewidth no longer follows a strict linear increase and instead tends toward a weaker $Q$-dependence. This deviation toward a plateau is a common indication that translational motion is not purely continuous on the corresponding molecular length scale, but involves finite residence periods between successive displacements. To account for this behavior, the data were fitted using the classical jump-diffusion expression

$$\Gamma_T(Q) = \frac{D_T Q^2}{1+\tau_0 D_T Q^2} \tag{4}$$

where $D_T$ is the translational diffusion coefficient and $\tau_0$ is the mean residence time between jumps.

Within this model, molecules are assumed to remain temporarily localized during a residence time $\tau_0$ and to undergo translational displacement through a sequence of discrete jumps. At low $Q$, the model reduces to the expected Fickian relation $\Gamma_T \approx D_T Q^2$, whereas at higher $Q$ the finite residence time limits the growth of the linewidth. The dashed-line fits shown in Fig. 6 describe the experimental $\Gamma_T(Q)$ satisfactorily for both the bulk and confined samples, supporting the view that the translational motion in the $LiCl.6H_2O$ aqueous solution on the IN13 timescale is well represented by a jump-diffusion process. A key result is that the linewidths of the confined sample are systematically smaller than those of the bulk at the same temperature and $Q$, which indicates slower dynamics under confinement.

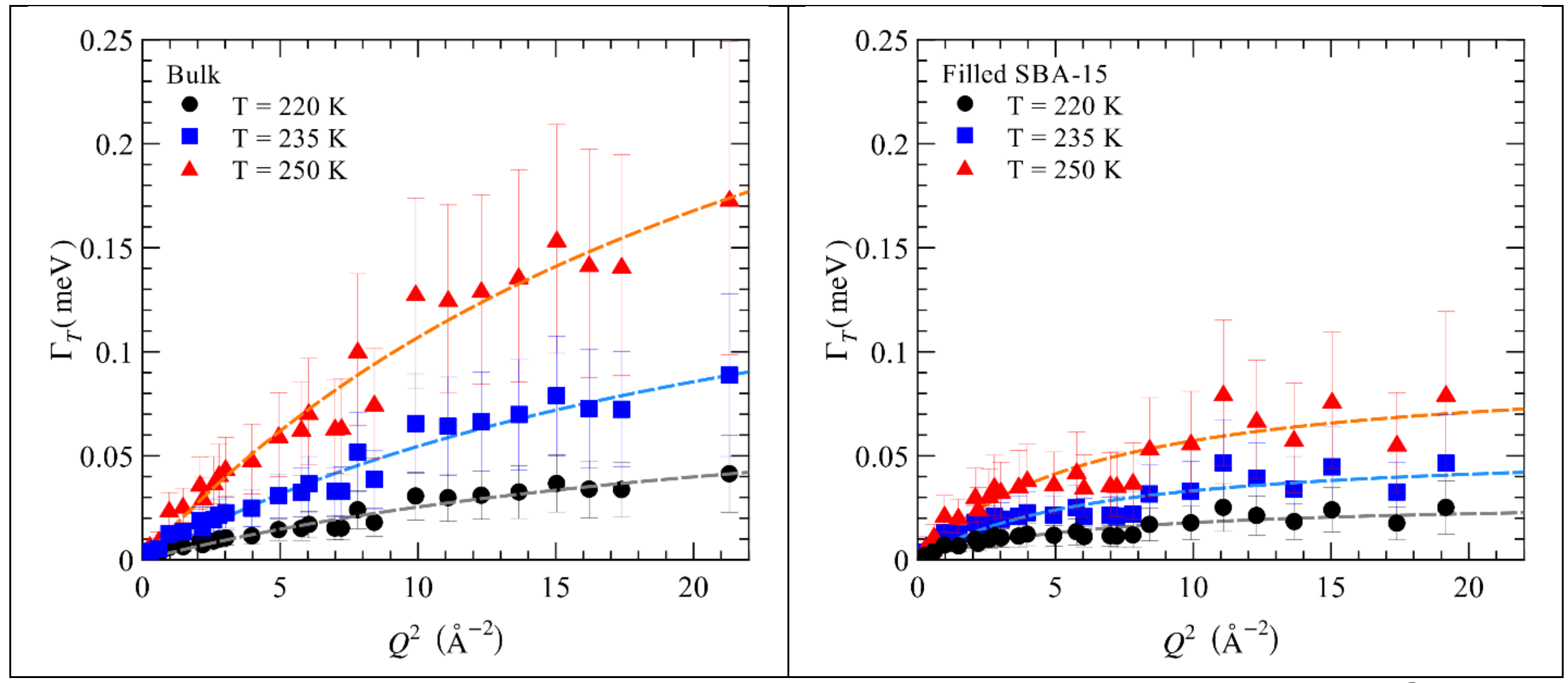

*Fig. 6. The linewidth $\Gamma_T$ derived from the IFWS fitting is plotted as a function of $Q^2$ for bulk $LiCl.6H_2O$ (left) and SBA-15-confined $LiCl.6H_2O$ (right) at selected temperatures. Dashed lines correspond to fits using the jump-diffusion model.*

The diffusion coefficient $D_T$ and the residence time $\tau_0$ are presented in an Arrhenius plot in Fig. 7. For comparison, we included QENS literature values from bulk and confined water,[35, 44, 48] as well as for bulk $LiCl.6H_2O$.[49] Our data for bulk $LiCl.6H_2O$ align well with earlier measurements obtained over a complementary temperature range.[49] However, some differences in the activation energy for $\tau_0$ values are observed. They can be attributed to variations in the analysis method, specifically Borreguero et al.[49] employed a stretched exponential fit rather than a simple Lorentzian. More importantly, the QENS instruments used also differ in their characteristics: IN13 covers a broader $Q$-range, which is particularly relevant for the determination of $\tau_0$.

Alternatively, one can extract a typical relaxation time defined by $\frac{1}{\tau_T(Q)} = \frac{D_T Q^2}{1+\tau_0 D_T Q^2}$ that can be compared to those obtained through different methods. The structural relaxation is often associated with the decay of $S(Q,t)$ at the main diffraction peak $Q_{max} \approx 2$Å$^{-1}$.[46] Although this definition is strictly valid for coherent rather than incoherent scattering, it provides a useful basis for comparison. In fact, the value of $\tau_T(Q_{max})$ shows reasonable agreement with relaxation times measured by other methods for LiCl aqueous solutions, including viscosity, dielectric spectroscopy, and NMR, as illustrated in supporting information.

First important conclusions concern the impact of confinement on the dynamics of LiCl.6$H_2O$. Compared with the bulk solution, the SBA-15-confined sample shows only a modest reduction in $D_T$, most clearly at the higher temperatures [Fig. 7(a)]. In contrast, the residence time is systematically larger in confinement than in the bulk [Fig. 7(b)], showing an increase of approximately 30%. This finding indicates that confinement mainly hinders translation by increasing the time spent in a transient local basin before a jump can occur. It points toward an increase in the lifetime of the cage formed by the neighboring molecules. Interestingly, our findings concerning the effect of confinement on the translational dynamics of LiCl.6$H_2O$ are similar to the observations made for pure water by Jani et al.[35, 44, 48] In the latter case, while the diffusion coefficient in SBA-15 shows only a slight reduction compared to bulk water, the residence time exhibits a significant increase (see dark red diamond in Fig. 7).

A second point of interest is the comparison between bulk LiCl.6$H_2O$ and bulk water. For this purpose, the present $D_T$ values were compared with literature data for bulk water.[35, 44, 48, 50] As expected, $D_T$ is clearly lower in the LiCl solution than in pure water, showing that the salt reduces long-range translational mobility. At the same time, however, $\tau_0$ is also lower in LiCl.6$H_2O$ than in bulk water. This apparently counterintuitive combination finds a logical interpretation: it shows that the salt does not simply slow water uniformly, but changes the microscopic diffusion mechanism itself. Mamontov demonstrated that, for concentrated LiCl solutions close to the present composition, water diffusion is slower than in pure water on long length scales.[51] Conversely, the observed reduction in residence time indicates that short-scale dynamics becomes more continuous, and so, less characteristic of the strongly associated jump-diffusion behavior typical of pure water. Borreguero and Mamontov later refined this interpretation, demonstrating that in LiCl·6$H_2O$, the salt disrupts water's hydrogen-bond network and inhibits the jump-diffusion process characteristic of pure water.[49] Their analysis suggests that the salt introduces shorter, more continuous translational movements, replacing the network-mediated motion of pure water. This results in a reduced residence time $\tau_0$, and a diffusion process that more closely resembles Fickian behavior, despite an overall decrease in the diffusion coefficient. Thus, the lower $D_T$ of bulk LiCl.6$H_2O$ reflects reduced long-range transport, whereas the lower $\tau_0$ reflects a salt-induced modification of the local diffusion mechanism, that is better described by the classical continuous Fickian diffusion due to disruption of the water hydrogen-bond network.

Coming back to the case of confined LiCl.6$H_2O$, the role of SBA-15 becomes more evident. In fact, the similar residence times observed for confined LiCl.6$H_2O$ and the bulk water are coincidental, arising from two opposing effects. The confined LiCl.6$H_2O$ should not be compared directly with pure water, but with a liquid whose microscopic dynamics has already been reshaped by strong ionic hydration. As a result, confinement acts on top of an already salt-modified dynamical state. Its primary effect is to increase $\tau_0$ again, showing that the pore environment reinforces local trapping without restoring the hydrogen-bond-network dynamics characteristic of pure water. Therefore, the confined liquid is best described as a salt-disrupted, hydration-controlled system whose water translational motion becomes further residence-limited in porous medium and most probably by pore-wall interactions.

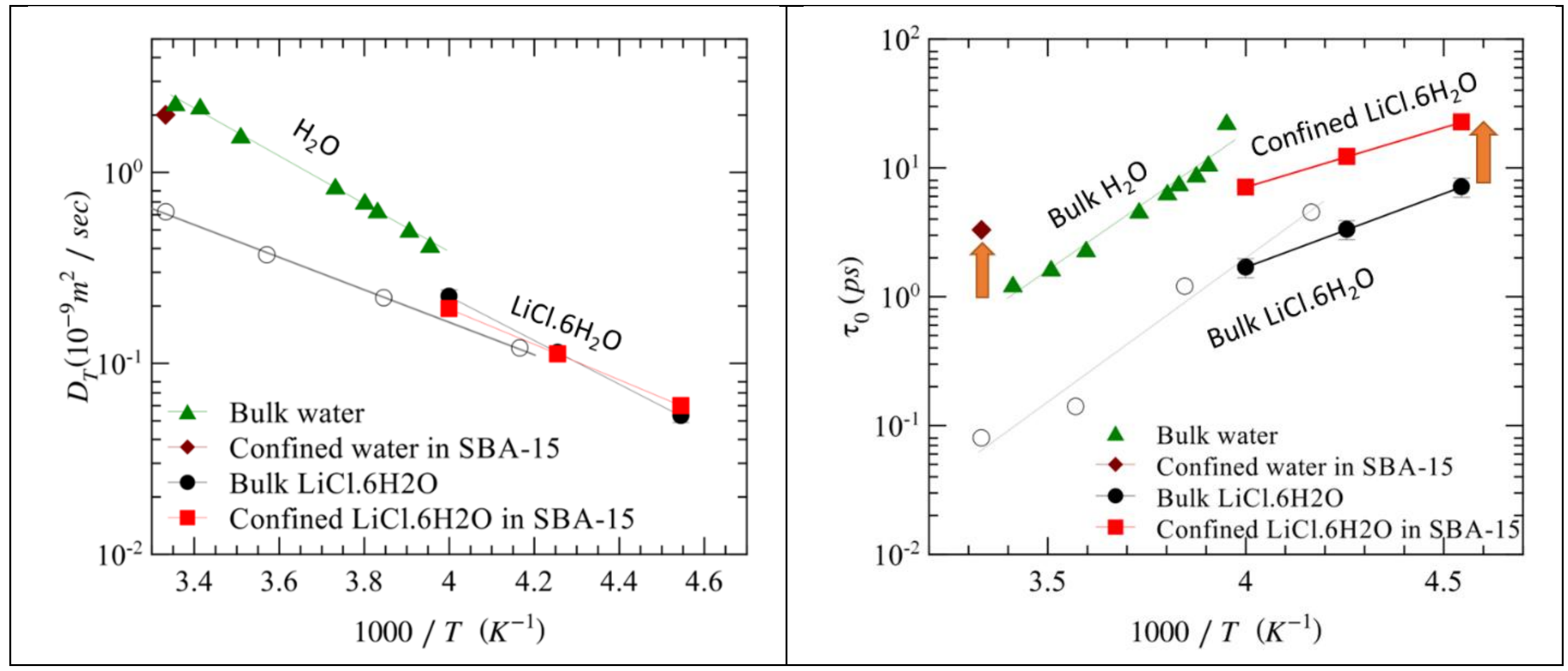

*Fig. 7. Temperature dependence of the effective translational diffusion coefficient, $D_T$ (left), and the residence time, $\tau_0$ (right), obtained from jump-diffusion fits of $\Gamma_T(Q)$ for bulk and SBA-15-confined LiCl.6$H_2O$. Filled black circles correspond to the present bulk LiCl.6$H_2O$ results, and filled red squares indicate the present SBA-15-confined LiCl.6$H_2O$ results. For comparison, literature values for bulk and confined water are shown as dark green triangles and dark red diamond, respectively (Ref. [35, 44, 48]), while the open symbols represent literature values for bulk LiCl.6$H_2O$ (Ref. [49]). Solid lines are guides for the eyes. The effect of confinement is illustrated by orange arrows in the right panel.*

This interpretation is supported by a direct inspection of the hydrogen-bond network using Raman spectra recorded in the O–H stretching region [Fig. 8]. In liquid water, this broad band is highly sensitive to hydrogen-bond strength and connectivity, with the lower-wavenumber side commonly associated with more strongly hydrogen-bonded environments and the higher-wavenumber side with weaker or less tetrahedrally coordinated O–H oscillators.[52-58] This makes Raman spectroscopy particularly useful for assessing how salt and confinement modify the local water structure.[38, 59-61] Compared with bulk water, bulk $LiCl.6H_2O$ exhibits a distinctly narrower spectral profile, along with a pronounced attenuation of the low-frequency shoulder near 3200 $cm^{-1}$, which is commonly associated with strongly hydrogen-bonded environments. This spectral redistribution indicates that the extended tetrahedral hydrogen-bond network of water is disrupted in the concentrated LiCl solution and replaced by a more loosely connected, hydration-controlled local structure. That interpretation is fully consistent with Raman studies of chloride salt solutions, including LiCl-containing systems, which show systematic reshaping of the O–H stretching band as ionic hydration competes with water–water hydrogen bonding.[60] Also, this Raman evidence supports the dynamical interpretation drawn from Fig. 7 by confirming that the hydrophilic confined medium does not restore the extensive hydrogen-bond network found in pure water.

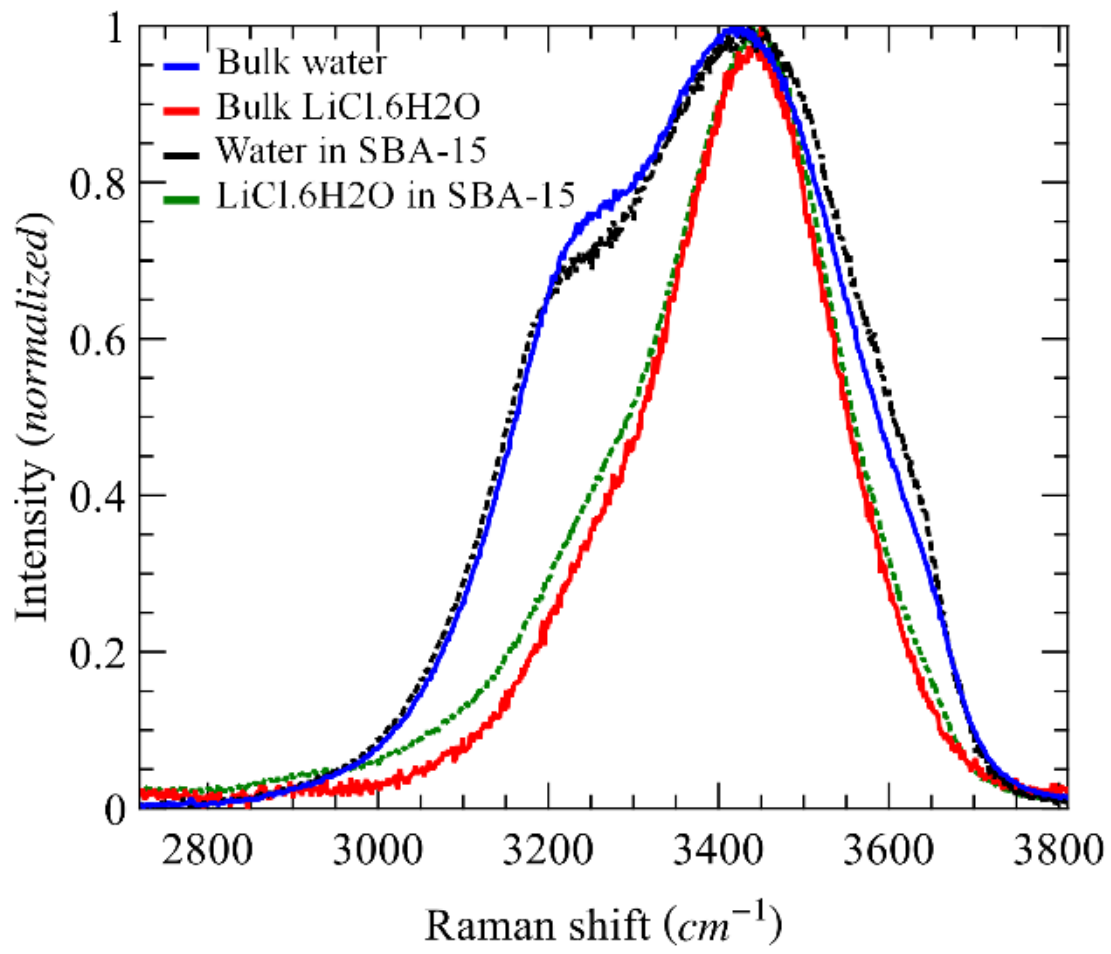

*Fig. 8. Normalized Raman spectra in the O–H stretching region for bulk water, bulk $LiCl.6H_2O$, water confined in SBA-15, and $LiCl.6H_2O$ confined in SBA-15.*

## 3.4. NMR measurements

### 3.4.1. Pulsed-field-gradient NMR

The fixed-window scan neutron analysis discussed above probes translational motion on the sub-nanosecond and nanometer scales and gives direct access to the local dynamics of bulk and $LiCl.6H_2O$ inside the pores. To place these results in a broader transport framework, it is useful to compare them with PFG-NMR self-diffusion measurements, which probe translational motion over much longer, micrometric length scales. Because these two techniques operate in very different dynamical windows, their comparison is therefore most meaningful at the level of physical trends rather than strict numerical equivalence, especially for the bulk sample. In this context, the bulk NMR behavior of $LiCl.6H_2O$ can be viewed in light of earlier studies on $LiCl.7H_2O$, which belongs to the same glass-forming concentration regime and exhibits coupled water–ion dynamics in the weakly supercooled state, followed by decoupling phenomena at lower temperature.[16, 23]

Fig. 9 compares the effective translational diffusion coefficients obtained from QENS with the self-diffusion coefficients measured by PFG-NMR for bulk $LiCl.6H_2O$ solution, together with literature data for bulk, confined water, and LiCl aqueous solution from the $^1H$ diffusion data of $LiCl.7H_2O$ from Vogel's research group.[21, 23, 48, 50] The confined $LiCl.6H_2O$ PFG-NMR data are not included here (the data are shown in Fig. S4), because their interpretation is less straightforward than for the bulk sample. For completeness, the corresponding semilogarithmic representation of the normalized attenuation is provided in the SI (Fig. S6). On the PFG-NMR length scale, the typical SBA-15 grain size and channel length, which are of the order of a few hundred nanometers. As a result, the measured attenuation is expected to contain contributions from restricted anisotropic pore diffusion, powder averaging over pore orientations, and possible exchange between different confinement environments, including intergrain and interparticle contributions. These coupled effects likely contribute to the very low apparent diffusion coefficients obtained for the confined sample, although their respective roles cannot be disentangled quantitatively from the present data. For this reason, the corresponding NMR-derived values are not treated here as unique self-diffusion coefficients of the confined liquid. In contrast, QENS probes translational motion on nanometer and sub-nanosecond scales and is therefore much more directly sensitive to the local dynamics of LiCl.6H2O inside the pores.

With this distinction in mind, Fig. 9 provides two clear messages. First, the reduced water mobility in bulk $LiCl.6H_2O$ solution compared to bulk water is observed for both methods (QENS and NMR) over the whole investigated temperature range, confirming that high salt concentration strongly suppresses translational motion. The bulk NMR literature data for $LiCl.7H_2O$ are in good overall agreement with the present bulk PFG-NMR results, which supports the view that $LiCl.6H_2O$ and $LiCl.7H_2O$ belong to the same glass-forming concentration regime and display comparable long-range transport behavior. Second, confinement further reduces the mobility, but the magnitude of this effect depends on the observation scale. On the neutron scale, the QENS-derived diffusion coefficients of confined $LiCl.6H_2O$ in 8 nm pores remain only moderately lower than those of the bulk solution. By contrast, the confined NMR literature data for $LiCl.7H_2O$ in 3.0 nm silica pores show a much stronger reduction relative to the bulk. This difference is physically reasonable because QENS probes local translational motion on nanometer and sub-nanosecond scales, whereas PFG-NMR reflects much longer displacements and is therefore more sensitive to restricted diffusion, pore geometry, and exchange between confinement environments. In this sense, Fig. 9 shows that confinement slows translational motion on both local and long-range scales, while affecting long-range transport more strongly than short-range mobility by considering the effect of the pore size. Similar observations were actually made for structured liquids, such as highly concentrated electrolytes[62] and deep eutectic mixtures.[37] Hence, these results indicate that the water dynamics in $LiCl.6H_2O$ remains active under confinement, but becomes progressively more hindered as the probed length scale increases. The comparison therefore supports the picture of a confined electrolyte whose local motion is retained to some extent, whereas its long-range translational transport is more strongly limited by the pore environment.

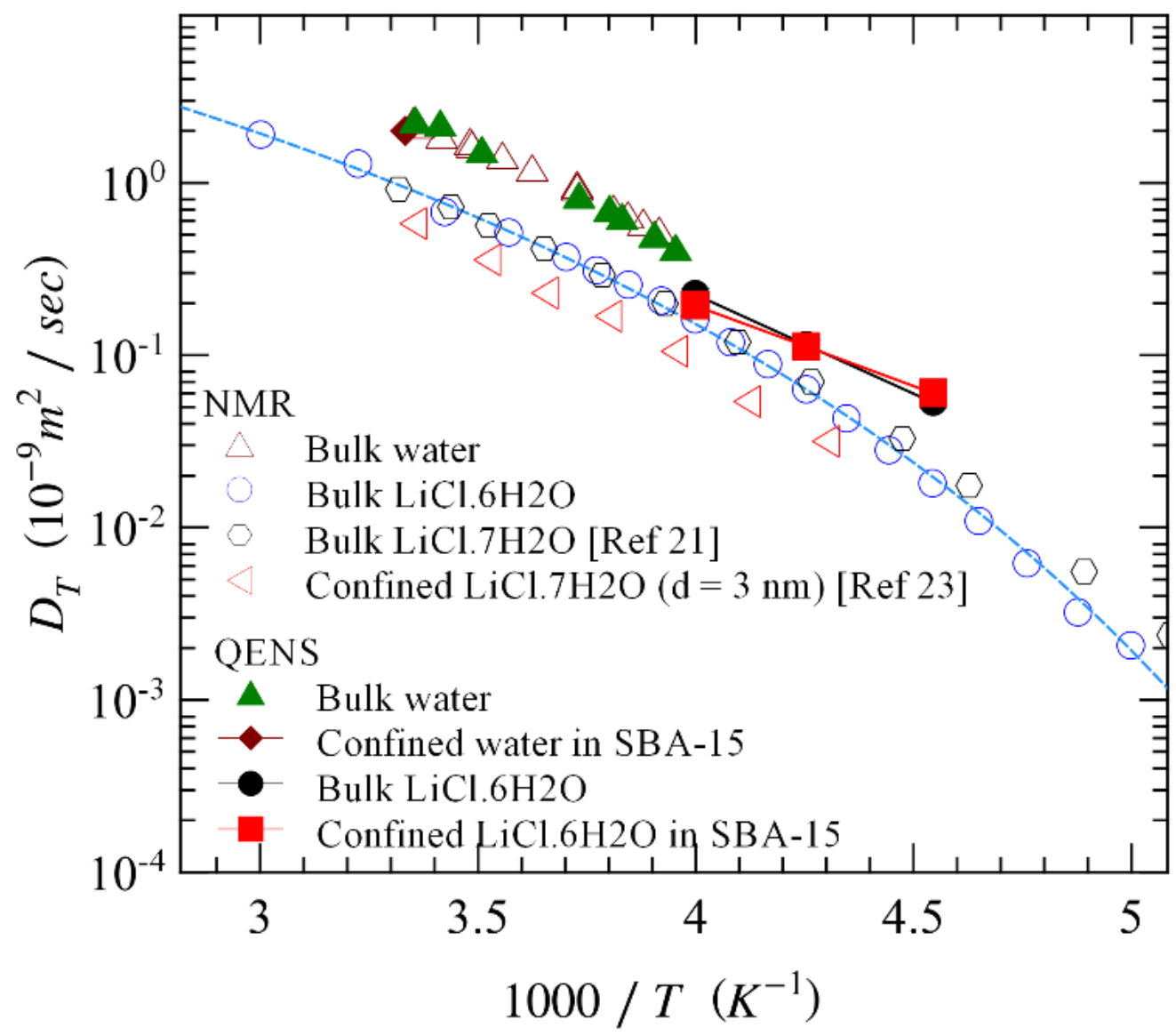

*Fig. 9. Arrhenius representation of the translational diffusion coefficient $D_T$ as a function of 1000/T for bulk water and LiCl.6H$_2$O solution. Filled symbols correspond to QENS-derived effective diffusion coefficients for bulk water, water confined in SBA-15, bulk LiCl.6H$_2$O, and LiCl.6H$_2$O confined in SBA-15. Open symbols correspond to PFG-NMR self-diffusion coefficients for bulk water and bulk LiCl.6H$_2$O, together with representative literature NMR data for bulk LiCl.7H$_2$O and LiCl.7H$_2$O confined in 3.0 nm silica pores reproduced from Schneider et al.,[21, 23] with the permission of AIP Publishing. Solid/dashed lines are guides to the eye.*

### 3.4.2. Spin–lattice relaxation

In addition to translational diffusion, the $^1$H spin–lattice relaxation time ($T_1$) provides access to the local proton dynamics of LiCl.6H$_2$O. In NMR, $T_1$ reflects the spectral density $J(\omega)$ of local magnetic-field fluctuations at the Larmor frequency. Within a simple single-correlation-time approximation, this dependence can be written in the familiar Bloembergen-Purcell-Pound expression as $\frac{1}{T_1} = C\left[\frac{\tau_c}{1+\omega_0^2\tau_c^2} + \frac{4\tau_c}{1+4\omega_0^2\tau_c^2}\right]$, where C is a coupling constant, $\tau_c$ is the correlation time, and $\omega_0$ is the Larmor angular frequency. In this framework, a $T_1$ minimum appears when the characteristic local correlation time becomes comparable to $1/\omega_0$. For the present $^1$H measurements at $\nu_0 = 300\ MHz$, this condition corresponds to

$$\tau_c^{min} = \frac{1}{2\pi\nu_0} = 0.53\ ns$$

Thus, the $T_1$ minimum observed here indicates that the relevant local proton fluctuations enter the sub-nanosecond time window of the present experiment. A useful bulk reference is provided by the $^1$H NMR study of Schneider and Vogel on LiCl.7H$_2$O, performed at a proton Larmor frequency of 91.2 MHz, where the $T_1$ minimum occurs around 100 ms at about 200 K and corresponds to a characteristic correlation time of about 1.7 ns.[21] The present minimum $T_1$ value is likewise close to that reported for LiCl.7H$_2$O and remains significantly higher than in pure water, suggesting that the intramolecular contribution dominates at such high salt concentrations. Since the $T_1$ minimum is reached when $\omega_0\tau_c$ is of order unity, that lower Larmor frequency corresponds to a characteristic timescale longer than in the present 300 MHz measurements. The comparison is physically meaningful because LiCl.7H$_2$O and LiCl.6H$_2$O belong to the same glass-forming concentration regime, but the different resonance frequencies must be taken into account when comparing the minimum positions and the associated correlation times. In the present case, $\tau_c^{min} \approx 0.53$ ns, which is consistent with probing a faster fluctuation window than in the earlier bulk $LiCl.7H_2O$ study.

As shown in Fig. 10, both bulk and SBA-15-confined $LiCl.6H_2O$ exhibit a clear $T_1$ minimum. This demonstrates that, in both systems, the local proton dynamics slows continuously on cooling and passes through the NMR observation window. So, the confined liquid remains dynamically active over the full investigated temperature range and cannot be described as a frozen interfacial phase. The main confinement effect is the broadening of the $T_1$ minimum together with lower $T_1$ values over a broad temperature interval, indicating that confinement broadens the distribution of local fluctuation times rather than simply shifting a single correlation time. This interpretation is consistent with the confinement studies of Schneider and co-workers, which showed that LiCl solutions in silica nanopores exhibit slowed but still mobile dynamics and, in sufficiently wide pores, may display distinct pore-center and wall-affected dynamical environments without evidence for a truly static Stern layer.[16, 23]

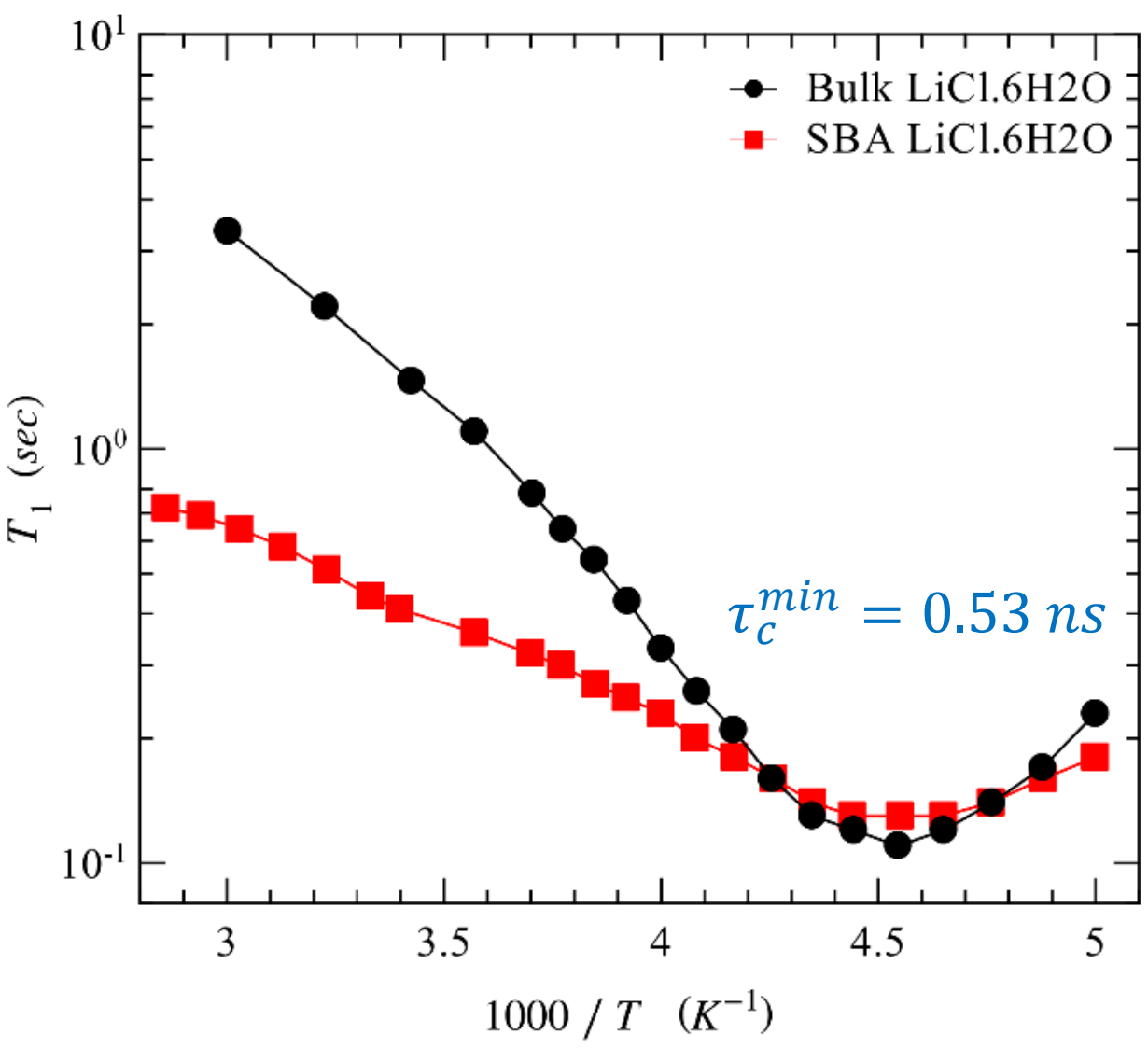

*Fig. 10. $T_1$ is plotted as a function of 1000/T for bulk $LiCl.6H_2O$ and $LiCl.6H_2O$ confined in SBA-15. Both samples exhibit a relaxation minimum, while the confined sample shows systematically shorter $T_1$ values over a broad temperature interval, indicating modified local correlation times and spectral density under confinement. The minimum correlation time is approximately 0.53 ns.*

This interpretation is consistent with the neutron results obtained here. The EFWS and IFWS analyses showed that confinement restricts the dynamical response and increases the residence time of translational motion on the neutron timescale. The proton $T_1$ data show that confinement also enhances the heterogeneous character of the local motions governing NMR relaxation. More generally, earlier NMR studies established that bulk LiCl aqueous solutions already exhibit complex multiscale dynamics, with different observables evolving differently upon deeper cooling.[16, 21] In this context, the broader $T_1$ minimum observed under confinement shows that SBA-15 superimposes an additional spatial distribution of local environments on an already hydration-controlled dynamical landscape. Overall, the $^1$H spin–lattice relaxation results indicate that LiCl.6$H_2O$ remains mobile in SBA-15, but with a broader distribution of local correlation times than in the bulk.

## Concluding remarks

The combined DSC, Raman, neutron-scattering, and NMR results show that confinement in SBA-15 preserves the glass-forming character of LiCl.6$H_2O$ while modifying the manner in which the dynamical slowdown develops across time and length scales. This behavior is consistent with earlier work establishing LiCl–$H_2O$ mixtures near eutectic composition as robust glass formers and showing that silica confinement alters their dynamics without producing a truly immobile liquid fraction.[5, 21, 27] Within the pore-size range investigated previously by NMR, the larger pores already showed that bulk-like pore-center dynamics can coexist with slower wall-affected motion, indicating that confinement acts by creating spatially distinct dynamical environments rather than by uniformly shifting the bulk behavior.[16, 23] Earlier NMR work also established that glass-forming LiCl aqueous solutions already show complex multiscale dynamics in the bulk.[21]

Against this background, the present results further show that salt and confinement contribute in fundamentally different ways. Raman spectroscopy confirms that in concentrated LiCl aqueous solution, the local water structure  is increasingly reorganized and governed by ion hydration and no longer dominated by the extended tetrahedral hydrogen-bond network of pure water, in agreement with earlier structural and spectroscopic studies of concentrated LiCl solutions.[17, 18, 38] Within this salt-modified liquid, neutron fixed-window analysis demonstrates that confinement reduces the motional amplitude on the IN13 timescale and broadens the temperature interval over which mobility develops, while the IFWS-based jump-diffusion analysis shows that translational motion becomes more residence-limited, as reflected by lower effective diffusivities and longer residence times. Together with the $^1$H relaxometry results, which reveal a broader distribution of local proton fluctuation times in confinement, these observations indicate that SBA-15 primarily enhances local trapping and spatial dynamical heterogeneity rather than restoring water-like dynamics or generating a frozen interfacial layer.[16, 23, 49, 51]

$LiCl.6H_2O$ confined within SBA-15 is therefore best described as a glass-forming electrolyte, where the salt disrupts the water structure and hydration dominates the dynamics. The intrinsic slowdown observed in bulk is further amplified in confinement by the porous media; however, the present data do not allow a unique separation of geometric restriction, interfacial water structuring, and ion–surface effects. In this sense, confinement does not simply retard the liquid uniformly. Instead, it reshapes the interplay between local mobility, translational jump motions, and dynamical heterogeneity by imposing a wider distribution of environments and a greater residence-time penalty on molecular motion.

## Supplementary Material

See the Supplementary Material for additional analysis and supporting data, including the extended treatment of the EFWS intensity using both Gaussian and generalized Zorn models, the corresponding temperature dependence of the mean-squared displacement for bulk and confined $LiCl.6H_2O$, representative IFWS fits at different momentum transfers for bulk and SBA-15-confined $LiCl.6H_2O$, and an Arrhenius comparison of the translational diffusion coefficients obtained by PFG-NMR and QENS for water and $LiCl.6H_2O$ in bulk and confinement.

## Acknowledgments

This research was carried out as part of the DFG-ANR SolutinPore collaborative project (Project AAPG2023), for which we express our gratitude. The authors gratefully acknowledge Dr. Marie-Vanessa Coulet and Killian Fourie (Aix Marseille Université, CNRS, MADIREL, Marseille) for performing the nitrogen sorption measurements used in the structural characterization of SBA-15. We also thank Claire Roiland (Institute of Chemistry, Rennes) for precious technical support on the PFG NMR experiments and the Institut Laue-Langevin for providing neutron beam time, which was essential to this work.

## Data availability

The raw data for this study were collected at the Institut Laue-Langevin (ILL) large-scale facility, and the dataset will be publicly available after the embargo period (i.e., five years after conducting the experiments) via the ILL Data Portal, using the provided DOI.[63] Data derived from these findings can also be obtained from the corresponding author upon request.

## References

(1) Demmel, F. Structural relaxation in the aqueous solution LiCl· 6D2O by quasielastic neutron scattering. *Journal of Molecular Liquids* **2021**, *332*, 115915. DOI: https://doi.org/10.1016/j.molliq.2021.115915.
(2) Corsaro, C.; Mallamace, D.; Cicero, N.; Vasi, S.; Dugo, G.; Mallamace, F. H NMR study of water clusters in supercooled LiCl/water solution. *Recent Advances on Energy, Environment, Ecosystems, and Development* **2015**.
(3) Mishima, O.; Stanley, H. E. The relationship between liquid, supercooled and glassy water. *Nature* **1998**, *396* (6709), 329–335. DOI: https://doi.org/10.1038/24540

(4) Achibat, T.; Duval, E.; Dupuy-Philon, J.; Jal, J.; Prevel, B.; Zorkani, I. Low-energy excitations in glass-forming aqueous lithium–chloride. *The Journal of chemical physics* **1995**, *102* (20), 8114–8117. DOI: https://doi.org/10.1063/1.469221.
(5) Longinotti, M. P.; Fuentes-Landete, V.; Loerting, T.; Corti, H. R. Glass transition of LiCl aqueous solutions confined in mesoporous silica. *The Journal of Chemical Physics* **2019**, *151* (6). DOI: https://doi.org/10.1063/1.5102142.
(6) Debenedetti, P. G.; Stanley, H. E. Supercooled and glassy water. *Physics Today* **2003**, *56* (6), 40–46. DOI: https://doi.org/10.1063/1.1595053.
(7) Angell, C. A. Insights into phases of liquid water from study of its unusual glass-forming properties. *Science* **2008**, *319* (5863), 582–587. DOI: https://doi.org/10.1126/science.1131939.
(8) Angell, C.; Sare, E.; Donnella, J.; MacFarlane, D. Homogeneous nucleation and glass transition temperatures in solutions of lithium salts in water-D2 and water. Doubly unstable glass regions. *The Journal of Physical Chemistry* **1981**, *85* (11), 1461–1464. DOI: https://doi.org/10.1021/j150611a001.
(9) Prevel, B.; Dupuy-Philon, J.; Jal, J.; Legrand, J.; Chieux, P. Structural relaxation in supercooled glass-forming solutions: a neutron spin-echo study of LiCl, 6D2O. *Journal of Physics: Condensed Matter* **1994**, *6* (7), 1279–1290. DOI: https://doi.org/10.1088/0953-8984/6/7/001.
(10) Suzuki, Y.; Mishima, O. Sudden switchover between the polyamorphic phase separation and the glass-to-liquid transition in glassy LiCl aqueous solutions. *The Journal of Chemical Physics* **2013**, *138* (8). DOI: https://doi.org/10.1063/1.4792498.
(11) Angell, C.; Sare, E. Liquid–Liquid Immiscibility in Common Aqueous Salt Solutions at Low Temperatures. *The Journal of Chemical Physics* **1968**, *49* (10), 4713–4714. DOI: https://doi.org/10.1063/1.1669935.

(12) Kobayashi, M.; Tanaka, H. Liquid-glass transition of water/salt mixtures. In *AIP Conference Proceedings*, 2013; American Institute of Physics: Vol. 1518, pp 252–259. DOI: https://doi.org/10.1063/1.4794576.
(13) Ruiz, G.; Bove, L. E.; Corti, H. R.; Loerting, T. Pressure-induced transformations in LiCl–H 2 O at 77 K. *Physical Chemistry Chemical Physics* **2014**, *16* (34), 18553–18562. DOI: https://doi.org/10.1039/C4CP01786B.
(14) Mishima, O. The glass-to-liquid transition of the emulsified high-density amorphous ice made by pressure-induced amorphization. *The Journal of chemical physics* **2004**, *121* (7), 3161–3164. DOI: https://doi.org/10.1063/1.1774151.
(15) Kobayashi, M.; Tanaka, H. Possible Link of the V-Shaped Phase Diagram to the Glass-Forming Ability and Fragility in a Water-Salt Mixture. *Physical review letters* **2011**, *106* (12), 125703. DOI: https://doi.org/10.1103/PhysRevLett.106.125703

(16) Schneider, S.; Brodrecht, M.; Breitzke, H.; Wissel, T.; Buntkowsky, G.; Varol, H.; Brilmayer, R.; Andrieu-Brunsen, A.; Vogel, M. Local and diffusive dynamics of LiCl aqueous solutions in pristine and modified silica nanopores. *The Journal of Chemical Physics* **2022**, *157* (3). DOI: https://doi.org/10.1063/5.0098483.
(17) Tromp, R.; Neilson, G.; Soper, A. Water structure in concentrated lithium chloride solutions. *The Journal of chemical physics* **1992**, *96* (11), 8460–8469. DOI: https://doi.org/10.1063/1.462298.
(18) Prével, B.; Jal, J.; Dupuy-Philon, J.; Soper, A. Structural characterization of an electrolytic aqueous solution, LiCl· 6H2O, in the glass, supercooled liquid, and liquid states. *The Journal of chemical physics* **1995**, *103* (5), 1886–1896. DOI: https://doi.org/10.1063/1.469713.
(19) Prevel, B.; Jal, J.; Dupuy-Philon, J.; Soper, A. Collective dynamical aspect related to vitrification in connection with structural characterization. *Physica A: Statistical Mechanics and its Applications* **1993**, *201* (1-3), 312–317. DOI: https://doi.org/10.1016/0378-4371(93)90428-7.
(20) Jal, J.; Soper, A.; Carmona, P.; Dupuy, J. Microscopic structure of LiCl. 6D2O in glassy and liquid phases. *Journal of Physics: Condensed Matter* **1991**, *3* (5), 551–567. DOI: https://doi.org/10.1088/0953-8984/3/5/005.
(21) Schneider, S.; Vogel, M. NMR studies on the coupling of ion and water dynamics on various time and length scales in glass-forming LiCl aqueous solutions. *The Journal of Chemical Physics* **2018**, *149* (10). DOI: https://doi.org/10.1063/1.5047825.
(22) Feiweier, T.; Isfort, O.; Geil, B.; Fujara, F.; Weingärtner, H. Decoupling of lithium and proton self-diffusion in supercooled LiCl: 7H2O: A nuclear magnetic resonance study using ultrahigh magnetic field gradients. *The Journal of chemical physics* **1996**, *105* (14), 5737–5744. DOI: https://doi.org/10.1063/1.472418.
(23) Schneider, S.; Säckel, C.; Brodrecht, M.; Breitzke, H.; Buntkowsky, G.; Vogel, M. NMR studies on the influence of silica confinements on local and diffusive dynamics in LiCl aqueous solutions approaching their glass transitions. *The Journal of chemical physics* **2020**, *153* (24). DOI: https://doi.org/10.1063/5.0036079.
(24) Khoder, H.; Siboulet, B.; Ollivier, J.; Baus-Lagarde, B.; Rébiscoul, D. How cation–silica surface interactions affect water dynamics in nanoconfined electrolyte solutions. *The Journal of Physical Chemistry C* **2024**, *128* (30), 12558–12565. DOI: https://doi.org/10.1021/acs.jpcc.4c02752.
(25) Martinez Casillas, D. C.; Longinotti, M. P.; Bruno, M. M.; Vaca Chavez, F.; Acosta, R. H.; Corti, H. R. Diffusion of water and electrolytes in mesoporous silica with a wide range of pore sizes. *The Journal of Physical Chemistry C* **2018**, *122* (6), 3638–3647. DOI: https://doi.org/10.1021/acs.jpcc.7b11555.
(26) Mamontov, E.; Cole, D. R.; Dai, S.; Pawel, M. D.; Liang, C.; Jenkins, T.; Gasparovic, G.; Kintzel, E. Dynamics of water in LiCl and CaCl2 aqueous solutions confined in silica matrices: A backscattering neutron spectroscopy study. *Chemical Physics* **2008**, *352* (1-3), 117–124. DOI: https://doi.org/10.1016/j.chemphys.2008.05.019.
(27) Jantsch, E.; Weinberger, C.; Tiemann, M.; Koop, T. Phase transitions of ice in aqueous salt solutions within nanometer-sized pores. *The Journal of Physical Chemistry C* **2019**, *123* (40), 24566–24574. DOI: https://doi.org/10.1021/acs.jpcc.9b06527.
(28) Meissner, J.; Prause, A.; Findenegg, G. H. Secondary confinement of water observed in eutectic melting of aqueous salt systems in nanopores. *The Journal of Physical Chemistry Letters* **2016**, *7* (10), 1816–1820. DOI: https://doi.org/10.1021/acs.jpclett.6b00756.
(29) Berlie, A.; Cavaye, H.; Peters, J. Exploring Molecular Motion through an Order–Disorder Transition within the Erythrosiderite Mineral Compound,(NH4) 2 [FeCl5 (H2O)]. *The Journal of Physical Chemistry C* **2024**, *129* (1), 447–452. DOI: https://doi.org/10.1021/acs.jpcc.4c05935.
(30) Morineau, D.; Xia, Y.; Alba-Simionesco, C. Finite-size and surface effects on the glass transition of liquid toluene confined in cylindrical mesopores. *The Journal of chemical physics* **2002**, *117* (19), 8966–8972. DOI: https://doi.org/10.1063/1.1514664.
(31) Alba-Simionesco, C.; Dosseh, G.; Dumont, E.; Frick, B.; Geil, B.; Morineau, D.; Teboul, V.; Xia, Y. Confinement of molecular liquids: Consequences on thermodynamic, static and dynamical properties of benzene and toluene. *The European Physical Journal E* **2003**, *12* (1), 19–28. DOI: https://doi.org/10.1140/epje/i2003-10055-1

(32) Morineau, D.; Guégan, R.; Xia, Y.; Alba-Simionesco, C. Structure of liquid and glassy methanol confined in cylindrical pores. *The Journal of chemical physics* **2004**, *121* (3), 1466–1473. DOI: https://doi.org/10.1063/1.1762872.
(33) Malfait, B.; Jani, A.; Morineau, D. Confining deep eutectic solvents in nanopores: Insight into thermodynamics and chemical activity. *Journal of Molecular Liquids* **2022**, *349*, 118488. DOI: https://doi.org/10.1016/j.molliq.2022.118488.
(34) Brodie-Linder, N.; Dosseh, G.; Alba-Simonesco, C.; Audonnet, F.; Impéror-Clerc, M. SBA-15 synthesis: Are there lasting effects of temperature change within the first 10 min of TEOS polymerization? *Materials chemistry and physics* **2008**, *108* (1), 73–81. DOI: https://doi.org/10.1016/j.matchemphys.2007.09.007.
(35) Jani, A.; Busch, M.; Mietner, J. B.; Ollivier, J.; Appel, M.; Frick, B.; Zanotti, J.-M.; Ghoufi, A.; Huber, P.; Fröba, M. Dynamics of water confined in mesopores with variable surface interaction. *The journal of chemical physics* **2021**, *154* (9). DOI: https://doi.org/10.1063/5.0040705.
(36) Jani, A.; Sohier, T.; Morineau, D. Phase behavior of aqueous solutions of ethaline deep eutectic solvent. *Journal of Molecular Liquids* **2020**, *304*, 112701. DOI: https://doi.org/10.1016/j.molliq.2020.112701.
(37) Kamar, M. N.; Mozhdehei, A.; Dupont, B.; Lefort, R.; Moréac, A.; Ollivier, J.; Appel, M.; Morineau, D. How special are the dynamics of deep eutectic solvents? A look at the prototypical case of ethaline. *The Journal of Chemical Physics* **2025**, *163* (13). DOI: https://doi.org/10.1063/5.0289812.
(38) Malfait, B.; Moréac, A.; Jani, A.; Lefort, R.; Huber, P.; Fröba, M.; Morineau, D. Structure of water at hydrophilic and hydrophobic interfaces: Raman spectroscopy of water confined in periodic mesoporous (organo) silicas. *The journal of physical chemistry C* **2022**, *126* (7), 3520–3531. DOI: https://doi.org/10.1021/acs.jpcc.1c09801.
(39) Mhanna, R.; Catrou, P.; Dutta, S.; Lefort, R.; Essafri, I.; Ghoufi, A.; Muthmann, M.; Zamponi, M.; Frick, B.; Morineau, D. Dynamic heterogeneities in liquid mixtures confined in nanopores. *The Journal of Physical Chemistry B* **2020**, *124* (15), 3152–3162. DOI: https://doi.org/10.1021/acs.jpcb.0c01035.
(40) Zorn, R. On the evaluation of neutron scattering elastic scan data. *Nuclear Instruments and Methods in Physics Research Section A: Accelerators, Spectrometers, Detectors and Associated Equipment* **2009**, *603* (3), 439–445. DOI: https://doi.org/10.1016/j.nima.2009.02.040.
(41) Klameth, F.; Vogel, M. Slow water dynamics near a glass transition or a solid interface: A common rationale. *The journal of physical chemistry letters* **2015**, *6* (21), 4385–4389. DOI: https://doi.org/10.1021/acs.jpclett.5b02010.
(42) Horstmann, R.; P Sanjon, E.; Drossel, B.; Vogel, M. Effects of confinement on supercooled tetrahedral liquids. *The Journal of Chemical Physics* **2019**, *150* (21). DOI: https://doi.org/10.1063/1.5095198.
(43) Frick, B.; Combet, J.; Van Eijck, L. New possibilities with inelastic fixed window scans and linear motor Doppler drives on high resolution neutron backscattering spectrometers. *Nuclear Instruments and Methods in Physics Research Section A: Accelerators, Spectrometers, Detectors and Associated Equipment* **2012**, *669*, 7–13. DOI: https://doi.org/10.1016/j.nima.2011.11.090.
(44) Teixeira, J.; Bellissent-Funel, M.-C.; Chen, S.-H.; Dianoux, A.-J. Experimental determination of the nature of diffusive motions of water molecules at low temperatures. *Physical Review A* **1985**, *31* (3), 1913. DOI: https://doi.org/10.1103/PhysRevA.31.1913.
(45) Qvist, J.; Schober, H.; Halle, B. Structural dynamics of supercooled water from quasielastic neutron scattering and molecular simulations. *The Journal of chemical physics* **2011**, *134* (14). DOI: https://doi.org/10.1063/1.3578472.
(46) Arbe, A.; Malo de Molina, P.; Alvarez, F.; Frick, B.; Colmenero, J. Dielectric susceptibility of liquid water: Microscopic insights from coherent and incoherent neutron scattering. *Physical review letters* **2016**, *117* (18), 185501. DOI: https://doi.org/10.1103/PhysRevLett.117.185501.
(47) Laage, D. Reinterpretation of the liquid water quasi-elastic neutron scattering spectra based on a nondiffusive jump reorientation mechanism. *The Journal of Physical Chemistry B* **2009**, *113* (9), 2684–2687. DOI: https://doi.org/10.1021/jp900307n.
(48) Price, W. S.; Ide, H.; Arata, Y. Self-diffusion of supercooled water to 238 K using PGSE NMR diffusion measurements. *The Journal of Physical Chemistry A* **1999**, *103* (4), 448–450. DOI: https://doi.org/10.1021/jp9839044.
(49) Borreguero, J. M.; Mamontov, E. Disruption of hydrogen-bonding network eliminates water anomalies normally observed on cooling to its calorimetric glass transition. *The Journal of Physical Chemistry B* **2017**, *121* (16), 4168–4173. DOI: https://doi.org/10.1021/acs.jpcb.7b01226.
(50) Baum, M.; Rieutord, F.; Juranyi, F.; Rey, C.; Rébiscoul, D. Dynamical and structural properties of water in silica nanoconfinement: impact of pore size, ion nature, and electrolyte concentration. *Langmuir* **2019**, *35* (33), 10780–10794. DOI: https://doi.org/10.1021/acs.langmuir.9b01434.
(51) Mamontov, E. Diffusion dynamics of water molecules in a LiCl solution: A low-temperature crossover. *The Journal of Physical Chemistry B* **2009**, *113* (43), 14073–14078. DOI: https://doi.org/10.1021/jp904734y.
(52) Hester, R. E. *AdVances in infrared and Raman spectroscopy*; Heyden, 1977.

(53) Du, Q.; Superfine, R.; Freysz, E.; Shen, Y. Vibrational spectroscopy of water at the vapor/water interface. *Physical Review Letters* **1993**, *70* (15), 2313. DOI: https://doi.org/10.1103/PhysRevLett.70.2313.
(54) Auer, B.; Kumar, R.; Schmidt, J.; Skinner, J. Hydrogen bonding and Raman, IR, and 2D-IR spectroscopy of dilute HOD in liquid D2O. *Proceedings of the National Academy of Sciences* **2007**, *104* (36), 14215–14220. DOI: https://doi.org/10.1073/pnas.0701482104.
(55) Brubach, J.-B.; Mermet, A.; Filabozzi, A.; Gerschel, A.; Roy, P. Signatures of the hydrogen bonding in the infrared bands of water. *The Journal of chemical physics* **2005**, *122* (18), 184509. DOI: https://doi.org/10.1063/1.1894929.
(56) Le Caër, S.; Pin, S.; Esnouf, S.; Raffy, Q.; Renault, J. P.; Brubach, J.-B.; Creff, G.; Roy, P. A trapped water network in nanoporous material: the role of interfaces. *Physical Chemistry Chemical Physics* **2011**, *13* (39), 17658–17666. DOI: https://doi.org/10.1039/C1CP21980D.
(57) Bergonzi, I.; Mercury, L.; Simon, P.; Jamme, F.; Shmulovich, K. Oversolubility in the microvicinity of solid–solution interfaces. *Physical Chemistry Chemical Physics* **2016**, *18* (22), 14874–14885. DOI: https://doi.org/10.1039/C5CP08012F.
(58) Mozhdehei, A.; Mercury, L.; Slodczyk, A. Ubiquity of the Micrometer-Thick Interface along a Quartz–Water Boundary. *Langmuir* **2024**, *40* (25), 13025–13041. DOI: https://doi.org/10.1021/acs.langmuir.4c00742.
(59) Li, R.; Jiang, Z.; Chen, F.; Yang, H.; Guan, Y. Hydrogen bonded structure of water and aqueous solutions of sodium halides: a Raman spectroscopic study. *Journal of molecular structure* **2004**, *707* (1-3), 83–88. DOI: https://doi.org/10.1016/j.molstruc.2004.07.016.
(60) Poulette, S. Raman Spectroscopic Properties of Aqueous Chloride Salt Solutions: Chlorides of Alkalis, Alkaline Earths and First-Row Transition Metals. **2022**. DOI: https://doi.org/10.7939/r3-mz0k-w587.
(61) Pamula, K.; Pophali, A.; Gersappe, D.; Kim, T. In-situ Raman spectroscopic analysis of phase transition temperatures in silica interacting with water and NaCl solutions. *Scientific Reports* **2025**, *15* (1), 26449. DOI: https://doi.org/10.1038/s41598-025-12034-2.
(62) Faraone, A.; Wang, F.; Takeuchi, S.; Dura, J. A. Structural relaxation and nanodomain dynamics in highly concentrated electrolytes for zinc batteries. *The Journal of Physical Chemistry C* **2024**, *128* (29), 12121–12130. DOI: https://doi.org/10.1021/acs.jpcc.4c00360.
(63) Morineau, D.; Belime, A.; Kamar, M.N.; Lefort, R.; Mozhdehei, A. Glassy Dynamics of LiCl Solutions in Nanoporous Media. **2025**. DOI: https://doi.ill.fr/10.5291/ILL-DATA.CRG-3268.